\documentclass{emulateapj}
\usepackage{amsmath,amssymb}
\usepackage{xspace}
\usepackage{graphicx}
\usepackage{epstopdf}
\usepackage{graphicx}
\usepackage{epsfig}
\usepackage{natbib}
\usepackage{CJK}

\def\beq{\begin{equation}}
\def\eeq{\end{equation}}
\def\mps{m~s$^{-1}$}
\def\msini{M\sin{i}}
\def\mjup{M_{\rm Jup}}
\def\msol{M_{\odot}}

\slugcomment{}
\shorttitle{}
\shortauthors{}

\begin{document}


\begin{CJK*}{UTF8}{gbsn}

\title
{
The Discovery of HD 37605\lowercase{$c$} and a Dispositive Null Detection of Transits of HD
37605\lowercase{$b$}\altaffilmark{1}
}

\altaffiltext{1}
{
Based on observations obtained with the Hobby-Eberly Telescope, which
is a joint project of the University of Texas at Austin, the
Pennsylvania State University, Stanford University, Ludwig Maximilians
Universit\"at M\"unchen, and Georg August Universit\"at G\"ottingen,
and observations obtained at the Keck Observatory, which is operated
by the University of California.  The Keck Observatory was made
possible by the generous financial support of the W. M. Keck
Foundation.
}

\author{Sharon Xuesong Wang (王雪凇)\altaffilmark{2,3},Jason T. Wright\altaffilmark{2,3}} 
\author{William Cochran\altaffilmark{4}} 
\author{Stephen R. Kane\altaffilmark{5}} 
\author{Gregory W. Henry\altaffilmark{6}} 
\author{Matthew J. Payne\altaffilmark{7}} 
\author{Michael Endl\altaffilmark{4}, Phillip J. MacQueen\altaffilmark{4}} 
\author{Jeff A. Valenti\altaffilmark{8}} 
\author{Victoria Antoci\altaffilmark{9,10}} 
\author{Diana Dragomir\altaffilmark{9}, Jaymie M. Matthews\altaffilmark{9}} 
\author{Andrew W. Howard\altaffilmark{11,12}} 
\author{Geoffrey W. Marcy\altaffilmark{11}} 
\author{Howard Isaacson\altaffilmark{11}} 
\author{Eric B. Ford\altaffilmark{7}} 
\author{Suvrath Mahadevan\altaffilmark{2,3}} 
\author{Kaspar von Braun\altaffilmark{5}} 

\altaffiltext{2}{Department of Astronomy and Astrophysics, 525 Davey
  Laboratory, The Pennsylvania State University, University Park, PA
  16802, USA; Send correspondence to xxw131@psu.edu and
  jtwright@astro.psu.edu}

\altaffiltext{3}{Center for Exoplanets and Habitable Worlds, 525 Davey
  Laboratory, The Pennsylvania State University, University Park, PA
  16802, USA}

\altaffiltext{4}{McDonald Observatory, The University of Texas,
  Austin, TX 78712, USA}

\altaffiltext{5}{NASA Exoplanet Science Institute, Caltech, MS 100-22,
  770 South Wilson Avenue, Pasadena, CA 91125, USA}

\altaffiltext{6}{Center of Excellence in Information Systems,
  Tennessee State University, 3500 John A. Merritt Blvd., Box 9501,
  Nashville,TN 37209, USA}

\altaffiltext{7}{Department of Astronomy, University of Florida, 211
  Bryant Space Science Center, P.O. Box 112055, Gainesville, FL 32611,
  USA}

\altaffiltext{8}{Space Telescope Science Institute, 3700 San Martin
  Dr., Baltimore, MD 21218, USA}

\altaffiltext{9}{Department of Physics \& Astronomy, University of
  British Columbia, Vancouver, BC V6T1Z1, Canada}

\altaffiltext{10}{Stellar Astrophysics Centre (SAC), Department of
  Physics and Astronomy, Aarhus University, Ny Munkegade 120, DK-8000
  Aarhus C, Denmark} 

\altaffiltext{11}{Department of Astronomy, University of California,
  Berkeley, CA 94720, USA}

\altaffiltext{12}{Space Sciences Laboratory, University of California,
  Berkeley, CA 94720, USA}

\begin{abstract}

We report the radial-velocity discovery of a second planetary mass
companion to the K0 V star HD 37605, which was already known to host
an eccentric, $P \sim 55$ days Jovian planet, HD 37605$b$. This second
planet, HD 37605$c$, has a period of $\sim 7.5$ years with a low
eccentricity and an $\msini$ of $\sim 3.4\ \mjup$. Our discovery was
made with the nearly 8 years of radial velocity follow-up at the
Hobby-Eberly Telescope and Keck Observatory, including observations
made as part of the Transit Ephemeris Refinement and Monitoring Survey
(TERMS) effort to provide precise ephemerides to long-period planets
for transit follow-up.
With a total of 137 radial velocity observations covering almost eight
years, we provide a good orbital solution of the HD 37605 system, and
a precise transit ephemeris for HD 37605$b$.
Our dynamic analysis reveals very minimal planet-planet interaction
and an insignificant transit time variation.
Using the predicted ephemeris, we performed a transit search for HD
37605$b$ with the photometric data taken by the T12 0.8-m Automatic
Photoelectric Telescope (APT) and the Microvariability and
Oscillations of Stars (MOST) satellite. Though the APT photometry did
not capture the transit window, it characterized the stellar activity
of HD~37605, which is consistent of it being an old, inactive star,
with a tentative rotation period of $57.67$ days. The MOST photometry
enabled us to report a dispositive null detection of a non-grazing
transit for this planet. Within the predicted transit window, we
exclude an edge-on predicted depth of 1.9\% at the $\gg 10\sigma$ level,
and exclude any transit with an impact parameter $b>0.951$ at greater
than $5\sigma$.
We present the BOOTTRAN package for calculating Keplerian orbital
parameter uncertainties via bootstrapping. We made a comparison and
found consistency between our orbital fit parameters calculated by the
RVLIN package and error bars by BOOTTRAN with those produced by a
Bayesian analysis using MCMC.

\end{abstract}


\section{Introduction}\label{sec:intro}

\subsection{Context}

Jupiter analogs orbiting other stars represent the first signposts of
true Solar System analogs, and the eccentricity distribution of these
planets with $a>3$ AU will reveal how rare or frequent true Jupiter
analogs are. To date, only 9 ``Jupiter analogs" have been
well-characterized in the peer reviewed literature\footnote{HD
  13931$b$ \citep{2010ApJ...721.1467H}, HD 72659$b$
  \citep{2011A&A...527A..63M}, 55 Cnc $d$ \citep{2002ApJ...581.1375M},
  HD 134987$c$ \citep{2010MNRAS.403.1703J}, HD 154345$b$ \citep[but
    with possibility of being an activity cycle-induced
    signal]{2008ApJ...683L..63W}, $\mu$ Ara $c$
  \citep{2007A&A...462..769P}, HD 183263$c$
  \citep{2009ApJ...693.1084W}, HD 187123$c$
  \citep{2009ApJ...693.1084W}, and GJ 832$b$
  \citep{2009ApJ...690..743B}.}  \citep[defined here as $P > 8$ years,
  $4 > M\sin{i} > 0.5\ \mjup$, and $e <
  0.3$;][exoplanets.org]{wrighteod}. As the duration of existing
planet searches approach 10--20 years, more and more Jupiter analogs
will emerge from their longest-observed targets
\citep{2012arXiv1205.2765W, 2012arXiv1205.5835B}.

Of the over 700 exoplanets discovered to date, nearly 200 are known to
transit their host star (\citealt{wrighteod}, exoplanets.org;
  \citealt{2011A&A...532A..79S}, exoplanet.eu), and many thousands more
candidates have been discovered by the {\it Kepler} telescope.  Of all
of these planets, only three orbit stars with $V<8$
\footnote{55 Cnc $e$ \citep{2004ApJ...614L..81M, 2011A&A...533A.114D},
  HD 189733 \citep{2005A&A...444L..15B}, and HD 209458
  \citep{2000ApJ...529L..41H, 2000ApJ...529L..45C}.}  and all have $P
< 4$ days.  Long period planets are less likely than close-in planets
to transit unless their orbits are highly eccentric and favorably
oriented, and indeed only 2 transiting planets with $P>20$ days have
been discovered around stars with $V<10$, and both have $e > 0.65$ (HD
  80606, \citealt{2009Natur.457..562L}, \citealt{2009MNRAS.396L..16F}; HD
  17156, \citealt{2007ApJ...669.1336F}, \citealt{2007A&A...476L..13B}; both
  highly eccentric systems were discovered first with radial
  velocities).

Long period planets not known to transit can have long transit windows
due to both the large duration of any edge-on transit and higher phase
uncertainties (since such uncertainties scale with the period of the
orbit). Long term radial velocity monitoring of stars, for instance
for the discovery of low amplitude signals, can produce collateral
benefits in the form of orbit refinement for a transit search and the
identification of Jupiter analogs
\citep[e.g.,][]{2009ApJ...693.1084W}. Herein, we describe an example
of both.

\subsection{Initial Discovery and Followup}

The inner planet in the system, HD 37605$b$, was the first planet
discovered with the Hobby-Eberly Telescope (HET) at McDonald
Observatory \citep{cochran2004}.  It is a super Jupiter ($\msini =
2.41\ \mjup$) on an eccentric orbit $e=0.67$ with an orbital period in
the ``period valley'' \citep[$P=55$ days;][]{2009ApJ...693.1084W}.

W.C., M.E., and P.J.M.\, of the University of Texas at
Austin, continued observations in order to get a much better orbit determination
and to begin searching for transits. With the first new data in the
fall of 2004, it became obvious that another perturber was present in
the system, first from a trend in the radial velocity (RV) residuals
\citep[i.e., a none-zero $\mbox{d} v/\mbox{d} t$; ][]{wit2007}, and
later from curvature in the residuals. By 2009, the residuals to a
one-planet fit were giving reasonable constraints on the orbit of a
second planet, HD 37605$c$, and by early 2011 the orbital parameters
of the $c$ component were clear, and the Texas team was preparing the
system for publication.

\subsection{TERMS Data}

The Transit Ephemeris Refinement and Monitoring Survey \citep[TERMS;][]{Kane2009} seeks
to refine the ephemerides of the known exoplanets orbiting bright,
nearby stars with sufficient precision to efficiently search for the
planetary transits of planets with periastron distances greater than a
few hundredths of an AU
\citep{2011EPJWC..1106005K, 2011ApJ...743..162P, 2011AJ....142..115D}.
This will provide the radii of planets not experiencing continuous high
levels of insolation around nearby, easily studied stars.

In 2010, S.M. and J.T.W.\ began radial velocity observations of HD
37605$b$ at HET from Penn State University for TERMS, to refine the
orbit of that planet for a future transit search. These observations,
combined with Keck radial velocities from the California Planet Survey (CPS)
consortium from 2006 onward, revealed that there was substantial
curvature to the radial velocity residuals to the original
\cite{cochran2004} solution. In October 2010 monitoring was
intensified at HET and at Keck Observatory by A.W.H., G.W.M., J.T.W.,
and H.I., and with these new RV data and the previously published
measurements from \cite{wit2007} they obtained a preliminary
solution for the outer planet. The discrepancy between the original
orbital fit and the new fit (assuming one planet) was presented at the
January 2011 meeting of the American Astronomical Society
\citep{terms2011aas}.

\subsection{Synthesis and Outline}

In early 2011, the Texas and TERMS teams combined efforts and began
joint radial velocity analysis, dynamical modeling, spectroscopic
analysis, and photometric observations \citep{terms2012aas}. The
resulting complete two-planet orbital solution allows for a
sufficiently precise transit ephemeris for the $b$ component to be
calculated for a thorough transit search. We herein report the transit
exclusion of HD 37605$b$ and a stable dynamical solution to the
system.

In \S~\ref{sec:spec}, we describe our spectroscopic observations and
analysis, which provided the radial velocities and the stellar
properties of HD 37605. \S~\ref{sec:orbit} details the orbital
solution for the HD 37605 system, including a comparison with MCMC
Keplerian fits, and our dynamical analysis. We report our photometric
observations on HD 37605 and the dispositive null detection\footnote{A
  {\em dispositive} null detection is one that disposes of the
  question of whether an effect is present, as opposed to one that
  merely fails to detect a purported or hypothetical effect that may
  yet lie beneath the detection threshold.  The paragon of dispositive
  null detections is the Michelson-Morley demonstration that the
  luminiferous ether does not exist \citep{michelson1887}.} of
non-grazing transits of HD 37605$b$ in \S~\ref{sec:photo}. After
\S~\ref{sec:summary}, Summary and Conclusion, we present updates on
$\msini$ of two previously published systems (HD 114762 and HD 168443)
in \S~\ref{sec:correction}. In the Appendix we describe the algorithm
used in the package BOOTTRAN (for calculating orbital parameter error
bars; see \S~\ref{sec:fit}).

\section{Spectroscopic Observations and Analysis}\label{sec:spec}
\subsection{HET and Keck Observations}
Observations on HD 37605 at HET started December of 2003. In total,
101 RV observations took place over the course of almost eight years,
taking advantage of the queue scheduling capabilities of HET.
The queue scheduling of HET allows for small amounts of telescope time
to be optimally used throughout the year, and for new observing
priorities to be implemented immediately, rather than on next
allocated night or after TAC and scheduling process \citep{hetque2007}.
The observations were taken through the High Resolution Spectrograph
\cite[HRS;][]{1998SPIE.3355..387T} situated at the basement of the HET
building. This
fiber-fed spectrograph has a typical long-term Doppler error of 3 -- 5
\mps\ \citep{2009MNRAS.393..969B}. The observations were taken with the
spectrograph configured at a resolving power of $R=$60,000. For more
details, see \cite{cochran2004}.

Observations at Keck were taken starting August 2006. A set of 33
observations spanning over five years were made through the HIRES
spectrometer \citep{1994SPIE.2198..362V} on the Keck I telescope,
which has a long-term Doppler error of 0.9 -- 1.5 \mps\
\citep[e.g.][]{2009ApJ...696...75H}. The observations were taken at a
resolving power of $R=$55,000. For more details, see
\cite{2009ApJ...696...75H} and \cite{2009ApJ...702..989V}.

Both our HET and Keck spectroscopic observations were taken with an
iodine cell placed in the light path to provide wavelength standard
and information on the instrument response function\footnote{Some
  authors refer to this as the ``point spread function'' or the
  ``instrumental profile'' of the spectrograph.}  (IRF) for radial
velocity extraction \citep{1992PASP..104..270M, 1996PASP..108..500B}.
In addition, we also have observations taken without iodine cell to
produce stellar spectrum templates -- on HET and Keck,
respectively. The stellar spectrum templates, after being deconvolved
with the IRF, are necessary for both radial velocity extraction and
stellar property analysis. The typical working wavelength range for
this technique is roughly 5000 \AA -- 6000 \AA.

\subsection{Data Reduction and Doppler Analysis}\label{sec:reduce}
In this section, we describe our data reduction and Doppler analysis
of the HET observations.  We reduced the Keck data with the standard
CPS pipeline, as described in, for example, \cite{2011ApJ...726...73H}
and \cite{2011ApJS..197...26J}.

We have constructed a complete pipeline for analyzing HET data -- from
raw data reduction to radial velocity extraction. The raw reduction is
done using the REDUCE package by \cite{2002A&A...385.1095P}. This
package is designed to optimally extract echelle spectra from 2-D
images \citep{1986PASP...98..609H}. Our pipeline corrects for cosmic
rays and scattered light.
In order to make the data reduction process completely automatic, we
have developed our own algorithm for tracing the echelle orders of HRS
and replaced the original semi-automatic algorithm from the REDUCE
package.

After the raw data reduction, the stellar spectrum template is
deconvolved using IRF derived from an iodine flat on the night of
observation. There were two deconvolved stellar spectrum templates
(DSST) derived from HET/HRS observations and one from
Keck/HIRES. Throughout this work, we use the Keck DSST, which is of
better quality thanks to a better known IRF of HIRES and a
superior deconvolution algorithm in the CPS pipeline
\citep{2009ApJ...696...75H, 2011ApJ...726...73H}.

Then the pipeline proceeds with barycentric correction and radial
velocity extraction for each observation. We have adopted the
Doppler code from CPS \citep[e.g.][]{
  2009ApJ...696...75H, 2011ApJ...726...73H, 2011ApJS..197...26J}. The
code is tailored to be fully functional with HET/HRS-formatted spectra,
and it is capable of working with either an HET DSST or a Keck one.

The 101 HET RV observations include 44 observations which
produced the published velocities in \cite{cochran2004} and
\cite{wit2007}, 34 observations also done by the Texas team in
follow-up work after 2007, and 23 observations taken as part of TERMS
program. We have performed re-reduction on these 44 observations
together with all the rest 57 HET observations through our
pipeline. This has the advantage of eliminating one free parameter in
the Keplerian fit -- the offset between two Doppler pipelines.

Two out of the 101 HET observations were excluded due to very low
average signal-to-noise ratio per pixel ($<20$), and one observation
taken at twilight was also rejected as such observation normally
results in low accuracy due to the significant contamination by the
residual solar spectrum (indeed this velocity has a residual of over
$100$ \mps\ against best Keplerian fit, much larger than the $\sim 8$
\mps\ RV error).

All the HET and Keck radial velocities used in this work (98 from HET
and 33 from Keck) are listed in Table \ref{rvtable}.

\subsection{Stellar Analysis}\label{sec:sme}

\renewcommand{\arraystretch}{1.2} 
\begin{deluxetable}{lc}
\tabletypesize{\scriptsize}
\tablecaption{STELLAR PARAMETERS\label{smetable}}
\tablewidth{180pt}
\tablehead{
  \colhead{Parameter} & \colhead{Value~~}
}
\startdata
~~Spectral type\tablenotemark{a} & K0 V ~~ \\
~~Distance (pc)\tablenotemark{a} & 44.0 $\pm$ 2.1 ~~ \\
~~$V$ & 8.661 $\pm$ 0.013 ~~ \\
~~$T_{\mbox{eff}}$ (K) & 5448 $\pm $44 ~~ \\
~~$\log{g}$ & 4.511 $\pm$ 0.024 ~~ \\
~~$[$Fe/H$]$ & 0.336 $\pm$ 0.030 ~~ \\
~~BC & -0.144 ~~ \\
~~$M_{\mbox{bol}}$ & 5.301 ~~ \\
~~$L_{\star}$ ($L_{\odot}$) & 0.590 $\pm$ 0.058 ~~ \\
~~$R_{\star}$ ($R_{\odot}$) & 0.901 $\pm$ 0.045\tablenotemark{c} ~~ \\ 
~~$M_{\star}$ ($\msol$) & 1.000 $\pm$ 0.050\tablenotemark{c} ~~ \\ 
~~$v\sin{i}$ & $<1$ k\mps ~~ \\
~~Age\tablenotemark{b}  & $\sim 7$ Gyr ~~
\enddata
\tablenotetext{a}{\cite{esa1997,vanleeuwen2008}.}
\tablenotetext{b}{\cite{svalue2010}, see \S~\ref{sec:apt}.}
\tablenotetext{c}{5\% relative errors, not the SME intrinsic
  errors. See footnote \ref{foot:sme} for details.}
\end{deluxetable}

HD 37605 is a K0 V star (V $\sim$ 8.7) with high proper motion at a
distance of 44.0 $\pm$ 2.1 pc \citep{esa1997,vanleeuwen2008}. We
derived its stellar properties based on analysis on a high-resolution
spectrum taken with Keck HIRES (without iodine cell in the light
path). Table \ref{smetable} lists the results of our
analysis\footnote{Note that the errors on the stellar radius $R_\star$
  and mass $M_\star$ listed in Table \ref{smetable} are not intrinsic to
  the SME code, but are 5\%$\times R_\star$ and 5\%$\times M_\star$. This is
  because the intrinsic errors reported by SME do not include the errors
  stemming from the adopted stellar models, and a more realistic
  precision for $R_\star$ and $M_\star$ would be around $\sim
  5$\%. Intrinsic errors reported by SME are $0.015\ L_\odot$ for
  $R_\star$ and $0.017\ \msol$ for $M_\star$.\label{foot:sme}}, including the
effective temperature $T_{\rm eff}$, surface gravity $\log{g}$, iron
abundance $[{\rm Fe/H}]$, projected rotational velocity $v\sin{i}$,
bolometric correction BC, bolometric magnitude $M_{\rm bol}$, stellar
luminosity $L_\star$, stellar radius $R_\star$, stellar mass $M_\star$
and age. HD 37605 is found to be a metal rich star ($[{\rm Fe/H}] \sim
0.34$) with $M_\star \sim 1.0\ \msol$ and $R_\star \sim 0.9
R_{\odot}$.

We followed the procedure described in \cite{valentifischer2005} and
also in \cite{2009ApJ...702..989V} with improvements. Briefly, the
observed spectrum is fitted with a synthetic spectrum using
Spectroscopy Made Easy \citep[SME; ][]{valentipiskunov1996} to derive
$T_{\rm eff}$, $\log{g}$, $[{\rm Fe/H}]$, $v\sin{i}$, and so on, which
are used to derive the bolometric correction BC and $L_\star$
consequently. Then an isochrone fit by interpolating tabulated
Yonsei-Yale isochrones \citep{2004ApJS..155..667D} using derived
stellar parameters from SME is performed to calculate $M_\star$ and
$\log{g_{\rm iso}}$ values (along with age and stellar radius). Next,
Valenti et al.~(2009) introduced an outside loop which re-runs SME
with $\log{g}$ fixed at $\log{g_{\rm iso}}$, followed by another
isochrone fit deriving a new log $\log{g_{\rm iso}}$ using the updated
SME results. The loop continues until $\log{g}$ values converge. This
additional iterative procedure to enforce self-consistency on log $g$
is shown to improve the accuracy of other derived stellar parameters
\citep{2009ApJ...702..989V}. The stellar radius and $\log{g}$ reported
here in Table \ref{smetable} are derived from the final isochrone fit,
which are consistent with the purely spectroscopic results. The
gravity ($\log{g} = 4.51$) is also consistent with the purely
spectroscopic gravity ($4.44$) based on strong Mg b damping wings, so
for HD 37605 the iteration process is optional.

\cite{cochran2004} reported the values of $T_{\rm eff}$, $\log{g}$,
and $[$Fe/H$]$ for HD 37605, and their estimates agree with ours
within 1$\sigma$ uncertainty. \cite{santos2005} also
estimated $T_{\rm eff}$, $\log{g}$, $[$Fe/H$]$, and $M_\star$, all of
which agree with our values within 1$\sigma$. Our stellar
mass and radius estimates are also consistent with the ones derived
from the empirical method by \cite{torres2010}.

Our SME analysis indicates that the rotation of the star ($v\sin{i}$)
is likely $< 1$ k\mps\ (corresponding to rotation period $\gtrsim 46$
days). We have used various methods to estimate stellar parameters
from the spectrum, including the incorporation of color and absolute
magnitude information and the Mg b triplet to constrain $\log{g}$, and
various macroturbulent velocity prescriptions.  All of these
approaches yield results consistent with an undetectable level of
rotational broadening, with an upper limit of 1-2 k\mps, consistent
with the tentative photometric period 57.67 days derived from the APT data (See
\S \ref{sec:apt}).

\section{Orbital Solution}\label{sec:orbit}

\begin{figure*}[th]
\epsscale{1.2}
\plotone{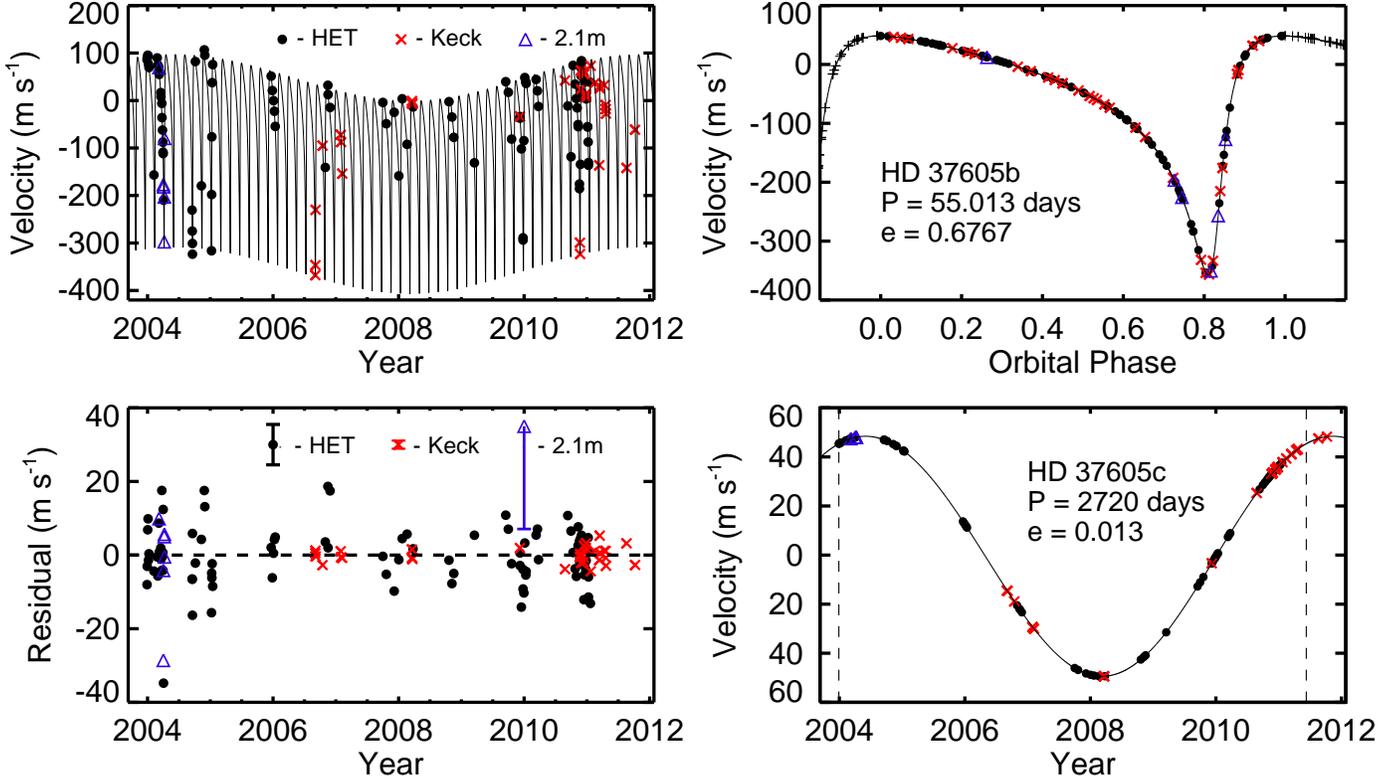}
\caption{Radial velocity and Keplerian model plots for the HD 37605
  system. In all panels, HET observations are labeled with black
  filled circles, Keck observations are labeled with red crosses, and
  the velocities from the 2.1 m telescope \citep{cochran2004} are
  labeled with blue triangles. Best Keplerian fits are plotted in
  black solid lines.
  {\bf Top left}: The best-fit 2-planet Keplerian model (solid
  line) and the observed radial velocities from 3 telescopes. The HET
  and Keck velocities have been adjusted to take into account the
  velocity offsets (i.e., subtracting $\Delta_{\rm HET}$ and
  $\Delta_{\rm Keck}$ from the velocities, respectively; see Table
  \ref{fittable} and \S~\ref{sec:fit}).
  {\bf Bottom left}: Residual velocities after subtracting the
  best-fit 2-planet Keplerian model. The lengend gives the typical
  size of the error bars using the $\pm$ median RV error for each
  telescope (for 2.1 m telescope only the lower half is shown).
  {\bf Top right}: RV signal induced by HD 37605$b$ alone, phased up
  to demonstrate our coverage. 
  {\bf Bottom right}: RV signal induced by HD 37605$c$ alone. The two
  vertical dashed lines denote the date of our first observation, and
  the date when HD 37605$c$ closes one orbit, respectively.
  (A color version of this figure is available in the online
  journal.) \label{rvplot}}
\end{figure*}

\subsection{Transit Ephemeris}\label{sec:emph}

The traditional parameters for reporting the ephemerides of
spectroscopic binaries are $P, K, e, \omega$, and $T_p$, the last
being the time of periastron passage \citep{2009ApJS..182..205W}.
This information is sufficient to predict the phase of a planet at any
point in the future in principle, but the uncertainties in those
parameters alone are insufficient to compute the uncertainty in
orbital phase without detailed knowledge of the covariances among the
parameters.

This problem is particularly acute when determining transit or
secondary eclipse times for planets with near circular orbits, where
$\sigma_{T_p}$ and $\sigma_\omega$ can be highly covariant.  In such
cases the circular case is often not excluded by the data, and so the
estimation of $e$ includes the case $e=0$, where $\omega$ is
undefined.  If the best or most likely value of $e$ in this case is
small but not zero, then it is associated with some nominal value of
$\omega$, but $\sigma_\omega$ will be very large (approaching $\pi$).
Since $T_p$ represents the epoch at which the true anomaly equals
$0$, $T_p$ will have a similarly large uncertainty (approaching
$P$), despite the fact that the phase of the system may actually be
quite precisely known!

In practice even the ephemerides of planets with well measured
eccentricities suffer from lack of knowledge of the covariance in
parameters, in particular $T_p$ and $P$ (whose covariance is sensitive
to the approximate epoch chosen for $T_p$).  To make matters worse,
the nature of ``1$\sigma$'' uncertainties in the literature is
inconsistent.  Some authors may report uncertainties generated while
holding all or some other parameters constant (for instance, by seeing
at what excursion from the nominal value $\chi^2$ is reduced by 1),
while others using bootstrapping or MCMC techniques may report the
variance in a parameter over the full distribution of trials.  In any
case, covariances are rarely reported, and in some cases authors even
report the most likely values on a parameter-by-parameter basis rather
than a representative ``best fit'', resulting in a set of parameters
that is not self-consistent.

The TERMS strategy for refining ephemerides therefore begins with the
recalculation of transit time uncertainties directly from the archival
radial velocity data.  We used bootstrapping (see Appendix) with the
time of conjunction, $T_c$ (equivalent to transit center, in the
case of transiting planets) computed independently for each trial.
For systems whose transit time uncertainty makes definitive
observations implausible or impossible due to the accumulation of
errors in phase with time, we sought additional RV measurements to
``lock down'' the phase of the planet.

\subsection{The 37605 System}\label{sec:fit}
There are in total 137 radial velocities used in the Keplerian fit for
the HD 37605 system. In addition to the 98 HET velocities and 33 Keck
ones (see \S \ref{sec:reduce}), we also included six\footnote{The
  velocity from observation on BJD $2,453,101.6647$ was rejected as it
  was from a twilight observation, which had both low precision
  ($\sigma_{\rm RV}=78.12$ \mps) and low accuracy (having a residual
  against the best Keplerian fit of over 100 \mps).} velocities from
\cite{cochran2004} which were derived from observations taken
with the McDonald Observatory 2.1 m Telescope (hereafter the 2.1 m
telescope).

We used the RVLIN package by \cite{2009ApJS..182..205W} to perform the
Keplerian fit. This package is based on the Levenberg\textendash
Marquardt algorithm and is made efficient in searching parameter space
by exploiting the linear parameters. The uncertainties of the
parameters are calculated through bootstrapping (with $1,000$
bootstrap replicates) using the BOOTTRAN package, which is described
in detail in the Appendix\footnote{The BOOTTRAN package is made
  publicly available online at http://exoplanets.org/code/ and the
  Astrophysics Source Code Library.}.

The best-fit Keplerian parameters are listed in Table
\ref{fittable}. The joint Keplerian fit for HD 37605$b$ and HD 37605$c$
has 13 free parameters: the orbital period $P$, time of periastron
passage $T_p$, velocity semi-amplitude $K$, eccentricity $e$, and the
argument of periastron referenced to the line of nodes $\omega$ for
each planet; and for the system, the velocity offset between the
center of the mass and barycenter of solar system $\gamma$ and two
velocity offsets between the three telescopes ($\Delta_{\rm Keck}$ and
$\Delta_{\rm HET}$, with respect to the velocities from the 2.1 m
telescope as published in \citealt{cochran2004}). We did not include
any stellar jitter or radial velocity trend in the fit (i.e., fixed to
zero). The radial velocity signals and the best Keplerian fits for the
system, HD 37605$b$ only, and HD 37605$c$ only are plotted in the
three panels of Fig.~\ref{rvplot}, respectively.

\begin{figure}[t]
\epsscale{1.15}
\plotone{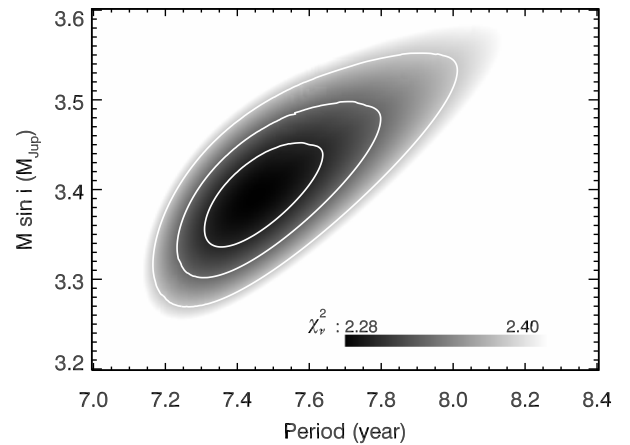} 
\caption{$\chi^2_{\nu}$ map for the best Keplerian fits with fixed
  values of period $P$ and minimum planet mass $\msini$ for HD
  37605$c$. This is showing that both $P$ and $\msini$ are
  well-constrained for this planet. The levels of the contours mark
  the 1$\sigma$ (68.27\%), 2$\sigma$ (95.45\%) and 3$\sigma$
  (99.73\%) confidence intervals for the 2-D $\chi^2$ distribution.\label{mmperplot}}
\end{figure}

Adopting a stellar mass of $M_\star=1.000 \pm 0.017\ \msol$ (as in Table
\ref{smetable}), we estimated the minimum mass ($\msini$) for HD
37605$b$ to be $2.802 \pm 0.011\ \mjup$ and $3.366 \pm 0.072\ \mjup$ for HD
37605$c$. While HD 37605$b$ is on a close-in orbit at $a=0.2831 \pm
0.0016$ AU that is highly eccentric ($e=0.6767 \pm 0.0019$), HD 37605$c$
is found to be on a nearly circular orbit ($e=0.013 \pm 0.015$) out at
$a=3.814 \pm 0.058$ AU, which qualifies it as one of the ``Jupiter
analogs''.

In order to see whether the period and mass of the outer planet, HD
37605$c$, are well constrained, we mapped out the $\chi_{\nu}^2$
values for the best Keplerian fit in the $P_c$-$M_c\sin{i}$ space
(subscript `$c$' denoting parameters for the outer planet, HD 37605$c$).
Each $\chi_{\nu}^2$ value on the $P_c$-$M_c\sin{i}$ grid was obtained
by searching for the best-fit model while fixing the period $P_c$ for the outer
planet and requiring constraints on $K_c$ and $e_c$ to maintain
$\msini$ fixed. As shown in Fig.~\ref{mmperplot}, our data are
sufficient to have both $P_c$ and $M_c\sin{i}$ well-constrained. This
is also consistent with the tight sampling distributions for $P_c$ and
$M_c\sin{i}$ found in our bootstrapping results.

The rms values against the best Keplerian fit are 7.86 \mps\ for HET,
2.08 \mps\ for Keck, and 12.85 \mps\ for the 2.1 m telescope. In the
case of HET and Keck, their rms values are slightly larger than their
typical reported RV errors ($\sim 5$ \mps\ and $\sim 1$ \mps,
respectively). This might be due to stellar jitter or underestimated
systematic errors in the velocities. We note that the $\chi_{\nu}^2$
is reduced to $1.0$ if we introduce a stellar jitter of 3.6 \mps\
(added in quadrature to all the RV errors).

\renewcommand{\arraystretch}{1.2} 
\begin{deluxetable}{lcc}
\tabletypesize{\scriptsize}
\tablecaption{KEPLERIAN FIT PARAMETERS\label{fittable}}
\tablewidth{240pt}
\tablehead{
  \colhead{Parameter} & \colhead{HD 37605$b$} & \colhead{HD 37605$c$}
}
\startdata
$P$ (days) & 55.01307 $\pm$ 0.00064 & 2720 $\pm$ 57 \\
{$T_p$ (BJD)\tablenotemark{a}} & 2453378.241 $\pm$ 0.020 & 2454838 $\pm$ 581 \\
{$T_c$ (BJD)\tablenotemark{b}} & 2455901.361 $\pm$ 0.069 & \nodata \\
$K$ (\mps) & 202.99 $\pm$ 0.72 & 48.90 $\pm$ 0.86 \\
$e$ & 0.6767 $\pm$ 0.0019 & 0.013 $\pm$ 0.015 \\
$\omega$ (deg) & 220.86 $\pm$ 0.28 & 221 $\pm$ 78 \\
$\msini$ ($\mjup$) & 2.802 $\pm$ 0.011 & 3.366 $\pm$ 0.072 \\
$a$ (AU) & 0.2831 $\pm$ 0.0016 & 3.814 $\pm$ 0.058 \\
$\gamma$ (\mps) & \multicolumn{2}{c}{$-$50.7 $\pm$ 4.6} \\
{$\Delta_{\rm Keck}$ (\mps)\tablenotemark{c}} &
                    \multicolumn{2}{c}{55.1 $\pm$ 4.7} \\
{$\Delta_{\rm HET}$ (\mps)\tablenotemark{c}} &
                    \multicolumn{2}{c}{36.7 $\pm$ 4.7} \\
$\chi^2_{\nu}$ & \multicolumn{2}{c}{2.28 ($d.o.f. = 124$)} \\
rms (\mps) & \multicolumn{2}{c}{7.61} \\
{Jitter (\mps)\tablenotemark{d}} & \multicolumn{2}{c}{3.6}
\enddata
\tablenotetext{a}{Time of Periastron passage.}
\tablenotetext{b}{Time of conjunction (mid-transit, if the system
  transits).}
\tablenotetext{c}{Offset with respect to the velocities from the 2.1 m
  telescope.}
\tablenotetext{d}{If a jitter of $3.6$ \mps\ is added in quadrature to
  all RV errors, $\chi^2_{\nu}$ becomes $1.0$.}
\end{deluxetable}

\begin{figure*}[!th]
\center
\includegraphics[angle=270,scale=1.0]{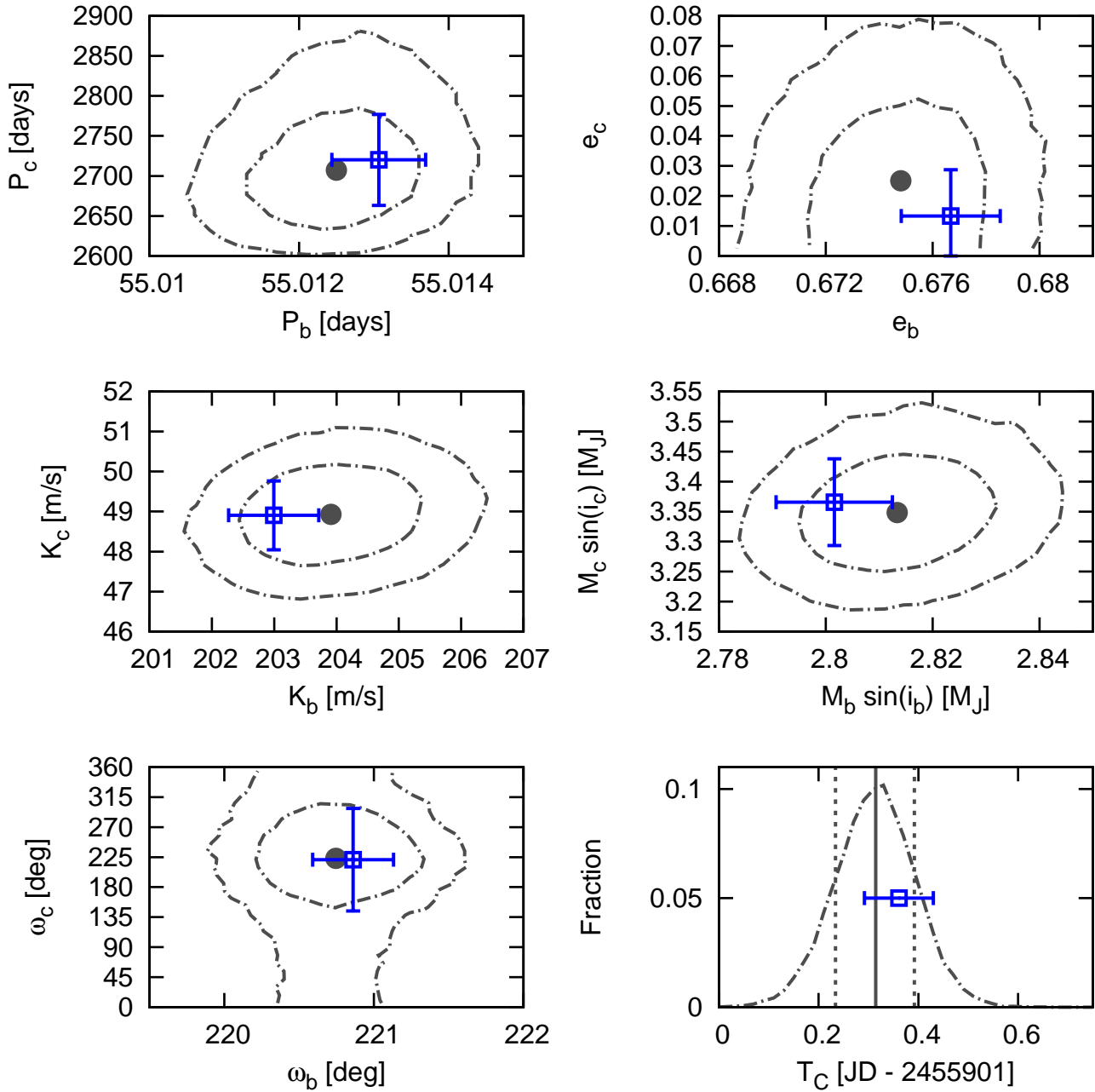}
\caption{Comparison between the Bayesian (MCMC) analysis and RVLIN$+$BOOTTRAN
  results.
  {\bf Top four and bottom left}: Contours of the posterior
  distributions of selected orbital parameters ($P$, $e$, $K$,
  $\msini$, and $\omega$) based on the MCMC analysis (dashed dotted
  line). The $x$-axes are orbital parameters of the inner planet, $b$,
  and the $y$-axes are those of the outer planet, $c$. The inner
  contours mark the 68.27\% (`1$\sigma$') 2-D confidence regions and
  the outer ones are 95.45\% (`2$\sigma$') ones. Also plotted are the
  best Keplerian fit from RVLIN (blue squares) and $\pm$1$\sigma$
  error bars estimated via bootstrapping (blue bars).
  {\bf Bottom right}: Marginalized posterior distribution of time of
  conjunction (mid-transit) $T_c$ of HD 37605$b$ in dashed dotted
  line. The solid grey vertical line is the median of the
  distribution, and the dashed grey vertical lines mark 1$\sigma$
  confidence interval. The blue square and its error bars are for the
  best estimate of $T_c$ from RVLIN$+$BOOTTRAN and its $\pm$1$\sigma$
  errors.
  See \S~\ref{SECN:MCMC} for details. 
  (A color version of this figure is available in the online
  journal.) \label{mcmcplot}}
\end{figure*}

\renewcommand{\arraystretch}{1.3} 
\begin{deluxetable*}{lllll}[!t]
\tabletypesize{\scriptsize}
\tablecaption{COMPARISON WITH MCMC RESULTS\label{mcmctable}}
\tablewidth{\textwidth}
\tablehead{
  \colhead{Parameter} & \multicolumn{2}{c}{HD 37605$b$} &
  \multicolumn{2}{c}{HD 37605$c$} \\
  \colhead{} & \colhead{RVLIN$+$BOOTTRAN} & \colhead{MCMC\tablenotemark{a}} &
  \colhead{RVLIN$+$BOOTTRAN} & \colhead{MCMC\tablenotemark{a}} 
}
\startdata
$P$ (days) & 55.01307 $\pm$ 0.00064 & $55.01250 \ {+ 0.00073} \ {-0.00075}$
           & 2720 $\pm$ 57 & $2707 \ {+57} \ {-42}$ \\

{$T_p$ (BJD)} & 2453378.243 $\pm$ 0.020 & $2453378.243 \ {+0.025} \ {-0.024}$
              & 2454838 $\pm$ 581 & $2454838 \ {+354} \ {-435}$ \\

{$T_c$ (BJD)} & 2455901.361 $\pm$ 0.069 & $2455901.314 \ {+0.077} \ {-0.081}$
              & \nodata & \nodata \\ 

$K$ (\mps) & 202.99 $\pm$ 0.72 & $203.91 \ {+0.92} \ {-0.88}$
           & 48.90 $\pm$ 0.86 & $48.93 \ {+0.82} \ {-0.82}$ \\

$e$ & 0.6767 $\pm$ 0.0019 & $0.6748 \ {+0.0022} \ {-0.0023}$
    & 0.013 $\pm$ 0.015  & $0.025 \ {+0.022} \ {-0.017}$ \\

$\omega$ (deg) & 220.86 $\pm$ 0.28 & $220.75 \ {+0.33} \ {-0.32}$
               & 221 $\pm$ 78 & $223 \ {+50} \ {-52}$ \\
M (deg)\tablenotemark{b} & 62.31 $\pm$ 0.15 & $62.27 \ {+0.18} \ {-0.18}$
          & 117 $\pm$ 78 & $118 \ {+56} \ {-51}$ \\

$\msini$ ($\mjup$) & 2.802 $\pm$ 0.011 & $2.814 \ {+0.012} \ {-0.012}$
                   & 3.366 $\pm$ 0.072 & $3.348 \ {+0.065} \ {-0.062}$  \\

$a$ (AU) & 0.2831 $\pm$ 0.0016 & $0.2833364 \ {+0.0000027} \ {-0.0000027}$
         & 3.814 $\pm$ 0.058 & $3.809 \ {+0.053} \ {-0.040}$  \\
Jitter (\mps)\tablenotemark{c} & 3.6  &  $2.70 \ {+0.53}
\ {-0.46}$ & & 
\enddata
\tablenotetext{a}{Median values of the marginalized posterior distributions and the
  68.27\% (`1$\sigma$') confidence intervals.}
\tablenotetext{b}{Mean anomaly of the first observation (BJD
  $2,453,002.671503$).}
\tablenotetext{c}{Like RVLIN, BOOTTRAN assumes no jitter or fixes jitter to a
  certain value, while MCMC treats it as a free parameter. See \S~\ref{SECN:MCMC}.} 
\end{deluxetable*}

\begin{figure*}[!th]
\center
\includegraphics[angle=270,scale=0.7]{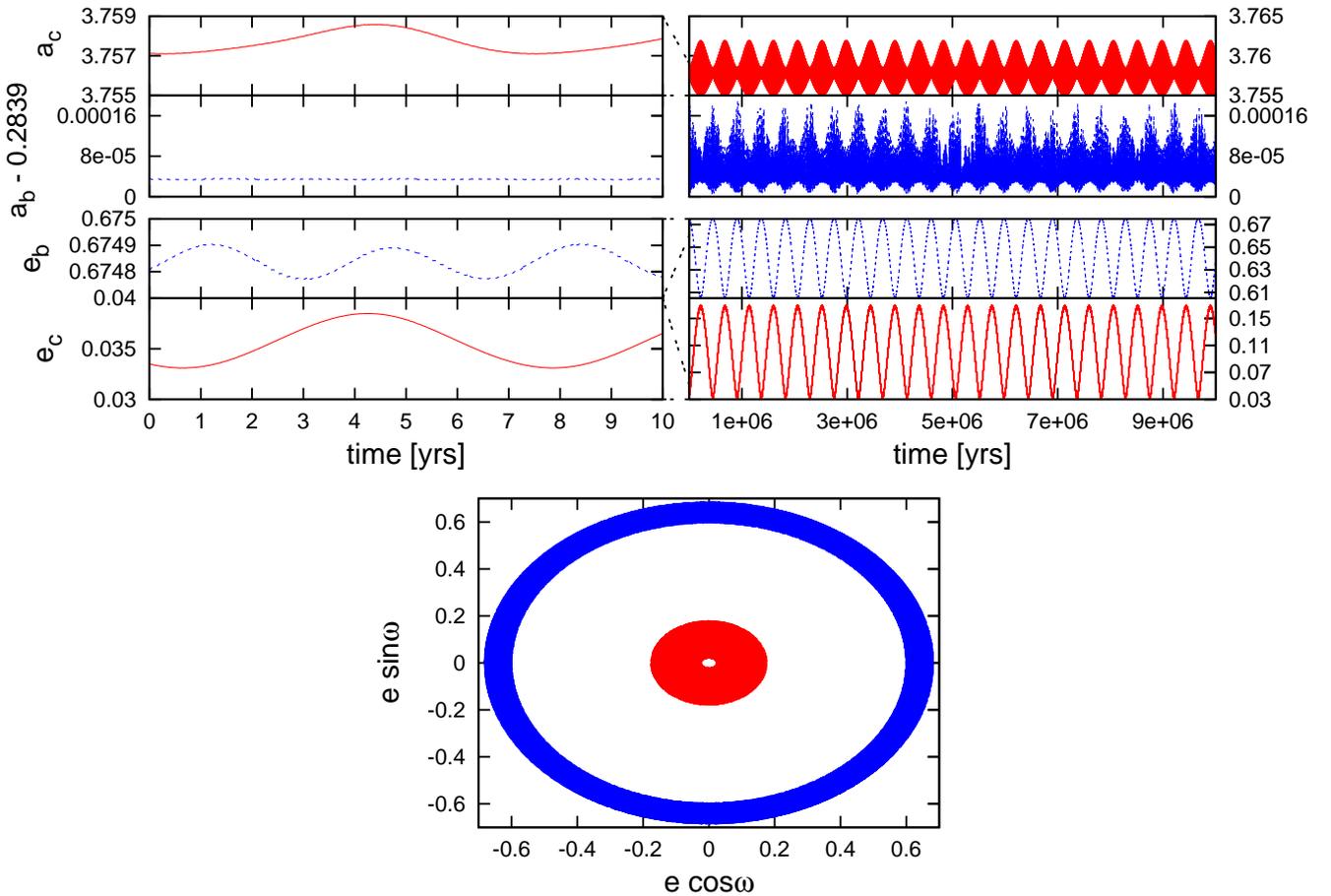}
\caption{Dynamic evolution of the best-fit MCMC system.
On the left we plot the short-term evolution over 10 years, on the
right we plot the evolution over $10^7$ years ($<1/10$ of our dynamic
simulation time scale).
The top plots describe the evolution of the semi-major axes and
eccentricities of the inner planet ($a_b$ \& $e_b$, blue dotted lines) and
the outer planet ($a_c$ \& $e_c$, red solid lines), while the bottom plot
describes the parameter space covered by the $e\cos{\omega},
e\sin{\omega}$ quantities over $10^8$ years (the blue and larger ring for
inner planet, and the red and smaller ring for outer planet).
We find that over the short-term (e.g., our RV observation window of
$\sim 10$ years), the parameter variations are negligible, but in the
long term significant eccentricity oscillations can take place
(particularly noticeable in the eccentricity of the outer planet).
See \S~\ref{SECN:DYNAMICS} for details.
(A color version of this figure is available in the online
journal.) \label{dynamicplot}}
\end{figure*}

\subsection{Comparison with MCMC Results}\label{SECN:MCMC}

We compared our best Keplerian fit from RVLIN and uncertainties
derived from BOOTTRAN (abbreviated as RVLIN$+$BOOTTRAN
hereafter) with that from a Bayesian framework following
\cite{Ford2005} and \cite{Ford2006} (referred to as the MCMC analysis
hereafter). Table \ref{mcmctable} lists the major orbital parameters
from both methods for a direct comparison. Fig.~\ref{mcmcplot}
illustrates this comparison, but with the MCMC results presented in
terms of 2-D confidence contours for $P$, $e$, $K$, $\msini$, and
$\omega$ of both planets, as well as for $T_c$ of HD 37605$b$.

For the Bayesian analysis, we assumed priors that are uniform in log
of orbital period, eccentricity, argument of pericenter, mean anomaly
at epoch, and the velocity zero-point.  For the velocity amplitude
($K$) and jitter ($\sigma_j$), we adopted a prior of the form
$p(x)=(x+x_o)^{-1}[log(1+x/x_o)]^{-1}$, with $K_o=\sigma_{j,o}=1$ \mps,
i.e. high values are penalized. For a detailed discussion of priors,
strategies to deal with correlated parameters, the choice of the
proposal transition probability distribution function, and other
details of the algorithm, we refer the reader to the original papers:
\cite{Ford2005,Ford2006,2007ASPC..371..189F}. The likelihood for
radial velocity terms assumes that each radial velocity observation
($v_i$) is independent and normally distributed about the true radial
velocity with a variance of $\sigma_i^2+\sigma_j^2$, where $\sigma_i$
is the published measurement uncertainty.  $\sigma_j$ is a jitter
parameter that accounts for additional scatter due to stellar
variability, instrumental errors and/or inaccuracies in the model
(i.e., neglecting planet-planet interactions or additional, low
amplitude planet signals).

We used an MCMC method based upon Keplerian orbits to calculate a
sample from the posterior distribution \citep{Ford2006}. We calculated
5 Markov chains, each with $\sim 2 \times 10^8$ states. We
discarded the first half of the chains and calculate Gelman-Rubin test
statistics for each model parameter and several ancillary
variables. We found no indications of non-convergence amongst the
individual chains. We randomly drew $3\times 10^4$ solutions from the
second half of the Markov chains, creating a sample set of the
converged overall posterior distribution of solutions. We then
interrogated this sample on a parameter-by-parameter basis to find the
median and $68.27\%\,(1$$\sigma)$ values reported in Table
\ref{mcmctable}. We refer to this solution set below as the
``best-fit'' MCMC solutions.

We note that the periods of the two planets found in this system are
very widely separated ($P_c/P_b \sim 50$), so we do \emph{not} expect
planet-planet interactions to be strong, hence we have chosen to forgo
a numerically intensive N-body DEMCMC fitting procedure \citep[see
  e.g.][]{Johnson2011,2011ApJ...729...98P} as the non-Keplerian
perturbations should be tiny (detail on the magnitude of the
perturbations is provided in \S \ref{SECN:DYNAMICS}).  However, to
ensure that the Keplerian fits generated are stable, we took the
results of the Keplerian MCMC fits and injected those systems into the
Mercury n-body package \citep{Chambers1999} and integrated them
forward for $\sim 10^8$ years. This allows us to verify that all of the
selected best-fit systems from the Keplerian MCMC analysis are indeed
long-term stable. Further details on the dynamical analysis of the
system can be found in \S \ref{SECN:DYNAMICS}.

We assumed that all systems are coplanar and edge-on for the sake of
this analysis, hence all of the masses used in our n-body analyses are
\emph{minimum} masses. 

As shown in Table \ref{mcmctable} and Fig.~\ref{mcmcplot}, the
parameter estimates from RVLIN$+$BOOTTRAN and MCMC methods agree with
each other very well (all within 1-$\sigma$ error bar). In some cases,
the MCMC analysis reports error bars slightly larger than
bootstrapping method ($\sim$ 20\% at most). We note that the
relatively large MCMC confidence intervals are \emph{not}
significantly reduced if one conducts an analysis at a \emph{fixed}
jitter level (e.g. $\sigma_J = 3.5$\mps) unless one goes to an
extremely low jitter value (e.g. $\sim 1.5$\mps). That is, the larger
MCMC error bars do not simply result from treating the jitter as a
free parameter.
For the uncertainties on minimum planet mass $\msini$ and semi-major
axes $a$, the MCMC analysis does not incorporate the errors on the
stellar mass estimate.
Note here, as previously mentioned in \S~\ref{sec:emph}, that
the ``best-fit" parameters reported by the MCMC analysis here listed
in Table \ref{mcmctable} are not a consistent set, as the best
estimates were evaluated on a parameter-by-parameter basis, taking the
median from marginalized posterior distribution of each. Assuming no
jitter, The best Keplerian fit from RVLIN has a reduced chi-square
value $\chi_{\nu}^2 = 2.28$, while the MCMC parameters listed in Table
\ref{mcmctable} give a higher $\chi_{\nu}^2$ value of 2.91.

~~
\subsection{Dynamical Analysis}\label{SECN:DYNAMICS}
%

We used the best-fit Keplerian MCMC parameters as the basis for a set
of long-term numerical (n-body) integrations of the HD 37605 system
using the Mercury integration package \citep{Chambers1999}. We used
these integrations to verify that the best-fit systems: (i) are
long-term stable; (ii) do not exhibit significant variations in their
orbital elements on the timescale of the observations (justifying the
assumption that the planet-planet interactions are negligible); (iii)
do not exhibit any other unusual features. We emphasize again that the
planets in this system are well separated and we do not expect any
instability to occur: for the masses and eccentricities in question, a
planet at $a_b \sim 0.28$ AU will have companion orbits which are
Hill stable for $a \gtrsim 0.83$ AU \citep{1993Icar..106..247G}, so while Hill
stability does not preclude outward scatter of the outer planet, the
fact that $a_c \sim 3.8 \gg 0.83$ AU suggests that the system will
be far from any such instability.

We integrated the systems for $> 10^8$ years ($\sim 10^7\times$ the
orbital period of the outer planet and $> 10^2\times$ the secular
period of the system), and plot in Fig. \ref{dynamicplot} the
evolution of the orbital elements $a,\ e$, \&\ $\omega$. On the
left-hand side of the plot we provide short-term detail, illustrating
that over the $\sim 10$ year time period of our observations, the
change in orbital elements will be very small.  On the right-hand side
we provide a much longer-term view, plotting $10^7$ out of $>10^8$ years
of system evolution, demonstrating that (i) the secular variation in
some of the elements (particularly the eccentricity of the outer
planet; see $e_c$ in red) over a time span of $\sim 4 \times 10^5$
years can be significant: in this case we see $0.03 < e_c < 0.11$, but
(ii) the system appears completely stable, as one would expect for
planets with a period ratio $P_c / P_b \sim 50$.  Finally, at the
bottom of the figure we display the range of parameter space covered
by the $e_i\cos{\omega_i},\ e_i\sin{\omega_i}$ parameters ($i=b$ in
blue for inner planet and $i=c$ in red for outer planet),
demonstrating that the orbital alignments circulate, i.e.\ they do not
show any signs of resonant confinement, which confirms our expectation
of minimal planet-planet interaction as mentioned before.

As noted above, our analysis assumed \emph{coplanar} planets. As such
the planetary masses used in these dynamical simulations are
\emph{minimum} masses. We note that for inclined systems, the larger
planetary masses will cause increased planet-planet perturbations. To
demonstrate this is still likely to be unimportant, we performed a
$10^8$ year simulation of a system in which $1 / \sin{i} = 10 $,
pushing the planetary masses to $\sim 30\ \mjup$. Even in such a
pathological system the eccentricity oscillations are only increased
by a factor of $\sim 2$ and the system remains completely stable for
the duration of the simulation.

We also performed a separate Transit Timing Variation (TTV) analysis,
using the best-fit MCMC systems as the basis for a set of highly
detailed short-term integrations. From these we extracted the times of
transit and found a TTV signal $\sim 100$ s, or $\sim 0.001$ day,
which is much smaller than the error bar on $T_c$ ($\sim 0.07$
day). Therefore we did not take into account the effect of TTV when
performing our transit analysis in the next section.

\subsection{Activity Cycles and Jupiter Analogs}\label{sec:activity}

The coincidence of the Solar activity cycle period of 11 years and
Jupiter's orbital period near 12 years illustrates how activity cycles
could, if they induced apparent line shifts in disk-integrated stellar
spectra, confound attempts to detect Jupiter analogs around Sun-like
stars.  Indeed, \cite{1985srv..conf..311D} predicted apparent radial
velocity variations of up to 30 \mps\ in solar lines due to the Solar
cycle, and \cite{1987ApJ...316..771D} reported a tentative detection
of such a signal in NIR CO lines of 30 \mps\ in just 2 years, and noted
that such an effect would severely hamper searches for Jupiter
analogs.  That concern was further amplified when
\cite{1991LNP...390...19C} reported a positive correlation between
radial velocity and chromospheric activity in the active star
$\kappa^1$ Cet, with variations of order 50--100 \mps.

\cite{2008ApJ...683L..63W} found that the star HD 154345 has an
apparent Jupiter analog (HD 154345 $b$), but that this star also shows
activity variations in phase with the radial velocity variations.
They noted that many Sun-like stars, including the precise radial
velocity standard star HD 185144 ($\sigma$ Dra) show similar activity
variations and that rarely, if ever, are these signals well-correlated
with signals similar in strength to that seen in HD 154345 ($\sim$ 15
\mps), and concluded that the similarity was therefore likely just an
inevitable coincidence.  Put succinctly, activity cycles in Sun-like
stars are common \citep{1995ApJ...438..269B}, but few Jupiter analogs
have been discovered, meaning that the early concern that activity
cycles would mimic giant planets is not a severe problem.

Nonetheless, there is growing evidence that activity cycles can, in
some stars, induce radial velocity variations
\citep{2010A&A...511A..54S,2011arXiv1107.5325L,dumusque2012},
and the example of HD 154345 still warrants care and concern. Most
significantly, \cite{2011A&A...535A..55D} found a positive correlation
between chromospheric activity and precise radial velocity in the
average measurements of a sample of HARPS stars, and provided a
formula for predicting the correlation strength as a function of the
metallicity and effective temperature of the star.  Their formulae
predict a value of 2 \mps\ for the most suspicious case in the
literature, HD 154345 (compared to an actual semiamplitude of $\sim$
15 \mps), but are rather uncertain.  It is possible that in a few,
rare cases, the formula might significantly underestimate the
amplitude of the effect.

\begin{figure}[!th]
\epsscale{1.15}
\plotone{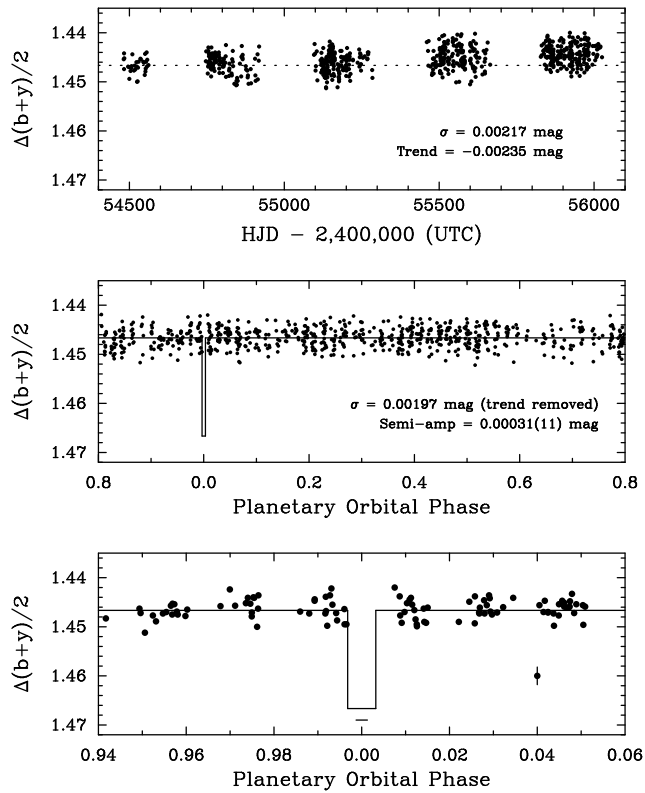}
\caption{Photometric observations of HD~37605 acquired over five years with 
 the T12 0.8m APT.  The top panel shows the entire five-year data set; the 
 dotted line represents the mean brightness of the first observing season.  A 
 long-term brightening trend is evident with a total range in the seasonal 
 means of 0.002 mag.  The middle panel shows the photometric data normalized 
 so that each season has the same mean as the first and then phased to the 
 orbital period of HD~37605$b$ (55.01307 day).  The solid line is the predicted 
 transit light curve, with Phase 0.0 being the predicted time of mid-transit, 
 $T_c$.  A least-squares sine fit of the phased data produces the very small 
 semi-amplitude of $0.00031~\pm~0.00011$ mag, providing strong evidence
 that the observed radial-velocity variations are not produced by
 rotational modulation of surface
 activity on the star.  The bottom panel plots the observations near $T_c$ 
 at an expanded scale on the abscissa.  The horizontal bar below the transit 
 window represents the $\pm$1$\sigma$ uncertainty in $T_c$.  Unfortunately, 
 none of the APT observations fall within the predicted transit window, so 
 we are unable to rule out transits with the APT observations.  See 
\S~\ref{sec:apt} for more. \label{photometry}}
\end{figure}

The top panel of Fig.~\ref{photometry} plots the T12 APT observations
from all five observing seasons (data provided in Table
\ref{phototable}; see details on APT photometry in
\S~\ref{sec:apt}). The dashed line marks the mean relative magnitude
($\Delta(b+y)/2$) of the first season.  The seasonal mean brightness
of the star increases gradually from year to year by a total of
$\sim 0.002$ mag, which may be due to a weak long-term magnetic cycle.
However, no evidence is found in support of such a cycle in the Mount
Wilson chromospheric Ca {\sc{ii}} H \& K indices \citep{svalue2010},
although the S values vary by approximately 0.1 over the span of a few
years. The formulae of \cite{2011arXiv1107.5325L} predict a
corresponding RV variation of less than 2 \mps\ due to activity, far too
small to confound our planet detection with $K=49$ \mps.

Since we do not have activity measurements for this target over the
span of the outer planet's orbit in HD 37605, we cannot definitively
rule out activity cycles as the origin of the effect, but the strength
of the outer planetary signal and the lack of such signals in other
stars known to cycle strongly dispels concerns that the longer signal
is not planetary in origin.

\section{The Dispositive Null Detection of Transits of HD 37605\lowercase{$b$}}\label{sec:photo}

\begin{figure}[!th]
\epsscale{1.15}
\plotone{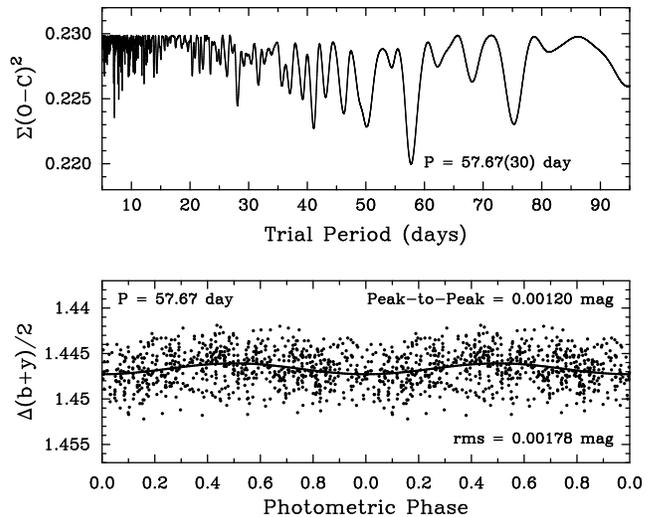}
\caption{Brightness variability in HD~37605 possibly induced by
  stellar rotation at $P = 57.67 \pm 0.30$ days.  Top panel is the
  periodogram of the complete, normalized data set.  Bottom panel
  shows the normalized photometry folded with this possible rotation
  period. The peak-to-peak amplitude is $0.00120~\pm~0.00021$ mag.
  See \S~\ref{sec:apt} for more.
\label{rotation}}
\end{figure}

We have performed a transit search for the inner planet of the system,
HD 37605$b$. This planet has a transit probability of 1.595\% and a
predicted transit duration of 0.352 day, as derived from the stellar 
parameters listed in Table \ref{smetable} and the orbital parameters given 
in Table \ref{fittable}. From the minimum planet mass ($\msini = 2.802
\pm 0.011\ \mjup$; see Table \ref{fittable}) and the models of
\cite{planetradius2003}, we estimate its radius to be $R_{\rm p}= 1.1\ R_{\rm Jup}$. 
Combined with the stellar radius of HD 37605 listed in Table~\ref{smetable}, 
$R_\star=0.901 \pm 0.015\ R_{\odot}$, we estimate the transit depth to be 
1.877\% (for an edge-on transit, $i=90^{\circ}$). We used both ground-based 
(APT; \S \ref{sec:apt}) and space-based (MOST; \S \ref{sec:most}) 
facilities in our search.

\subsection{APT Observations and Analysis}\label{sec:apt}

The T12 0.8-m Automatic Photoelectric Telescope (APT), located at Fairborn 
Observatory in southern Arizona, acquired 696 photometric observations of
HD~37605 between 2008 January 16 and 2012 April 7. \cite{henry1999} provides 
detailed descriptions of observing and data reduction procedures with 
the APTs at Fairborn. The measurements reported here are differential 
magnitudes in $\Delta (b+y)/2$, the mean of the differential magnitudes 
acquired simultaneously in the Str\"omgren $b$ and $y$ bands with two
separate EMI 9124QB bi-alkali photomultiplier tubes.
The differential magnitudes are computed from the mean of three 
comparison stars:
HD~39374 (V = 6.90, B$-$V = 0.996, K0 III), 
HD~38145 (V = 7.89, B$-$V = 0.326, F0 V), 
and HD~38779 (V = 7.08, B$-$V = 0.413, F4 IV).
This improves the precision of each individual measurement and helps to 
compensate for any real microvariability in the comp stars. Intercomparison 
of the differential magnitudes of these three comp stars demonstrates that all 
three are constant to 0.002 mag or better from night to night, consistent with
typical single-measurement precision of the APT 
\citep[0.0015-0.002 mag;][]{henry1999}.

\renewcommand{\arraystretch}{1.2}
\begin{deluxetable}{cc}
\tabletypesize{\small}
\tablewidth{180pt}
\tablecaption{PHOTOMETRIC OBSERVATIONS OF HD~37605 FROM THE T12 0.8m APT
\label{phototable}}
\tablehead{
\colhead{Heliocentric Julian Date} & \colhead{$\Delta (b+y)/2$} \\
\colhead{(HJD $-$ 2,400,000)} & \colhead{(mag)}
}
\startdata
54,481.7133 & 1.4454 \\
54,482.6693 & 1.4474 \\
54,482.7561 & 1.4442 \\
54,483.6638 & 1.4452 \\
54,495.7764 & 1.4469 \\
54,498.7472 & 1.4470 
\enddata
\tablecomments{This table is presented in its entirety in the electronic edition
of the Astrophysical Journal.  A portion is shown here for guidance regarding
its form and content.}
\end{deluxetable}


Fig.~\ref{photometry} illustrates the APT photometric data and our
transit search. As mentioned in \S~\ref{sec:activity}, the top panel
shows all of our APT photometry covering five observing seasons, which
exhibits a small increasing trend in the stellar brightness.
To search for the transit signal of HD 37605$b$, the photometric data
were normalized so that all five seasons had the same mean
(referred to as the ``normalized photometry" hereafter). The data were
then phased at the orbital period of HD~37605$b$, $55.01307$ days, and
the predicted time of mid-transit, $T_c$, defined as Phase~0.  The
normalized and phased data are plotted in the middle panel of
Fig.~\ref{photometry}.  The solid line is the predicted transit light
curve, with the predicted transit duration (0.352 day or 0.0064 phase
unit) and transit depth (1.877\% or $\sim 0.020$ mag) as estimated
above. The scatter of the phased data from their mean is $0.00197$
mag, consistent with APT's single-measurement precision, and thus
demonstrates that the combination of our photometric precision and the
stability of HD~37605 is easily sufficient to detect the transits of
HD~37605b in our phased data set covering five years.  A least-squares
sine fit of the phased data gives a very small semi-amplitude of
$0.00031~\pm~0.00011$ mag (consistent with zero) and so provides
strong evidence that the observed radial-velocity variations are not
produced by rotational modulation of surface activity on the star.

The bottom panel of Fig.~\ref{photometry} plots the phased data around
the predicted time of mid-transit, $T_c$, at an expanded scale on the
abscissa.  The horizontal bar below the transit window represents the
$\pm$1$\sigma$ uncertainty on $T_c$ (0.138 day or 0.0025 phase unit for 
$T_c$'s near BJD $2,455,901.361$; see \S~\ref{sec:fit}).  The light curve 
appears to be highly clustered, or binned, due to the near integral orbital
period ($P \sim 55.01$ days) and consequent incomplete sampling from a 
single observing site.  Unfortunately, none of the data clusters chance 
to fall within the predicted transit window, so we are unable to rule out 
transits of HD~37605b with the APT observations.

Periodogram analysis of the five individual observing seasons revealed no
significant periodicity between 1 and 100 days.  This suggests that the star 
is inactive and the observed $K \sim 200$ \mps\ RV signal (for HD~37605$b$) 
is unlikely to be the result of stellar activity.

Analysis of the complete, normalized data set, however, suggests a week 
periodicity of $57.67 \pm 0.30$ days with a peak-to-peak amplitude of just 
$0.0012 \pm 0.0002$ mag (see Fig.~\ref{rotation}).  We tentatively 
identify this as the stellar rotation period.  This period is consistent 
with the projected rotational velocity of $v\sin{i} < 1$ k\mps derived 
from our stellar analysis described in \S \ref{sec:sme}.  It is also 
consistent with the analysis of \cite{svalue2010}, who derived a Mount Wilson
chromospheric Ca {\sc{ii}} H \& K index of $S=0.165$, corresponding to
$\log R^\prime_{\rm HK} = -5.03$.  Together, these results imply a
rotation period $\gtrsim 46$ days and an age of $\sim 7$ Gyr (see 
Table~\ref{smetable}).  Similarly, \cite{age2002} find an age of $> 10$ Gyr 
using isochrones along with the Hipparcos parallax and space motion, 
supporting HD~37605's low activity and long rotation period.

\subsection{MOST Observations and Analysis}\label{sec:most}

\begin{figure}[!th]
\includegraphics[angle=270.,scale=0.37]{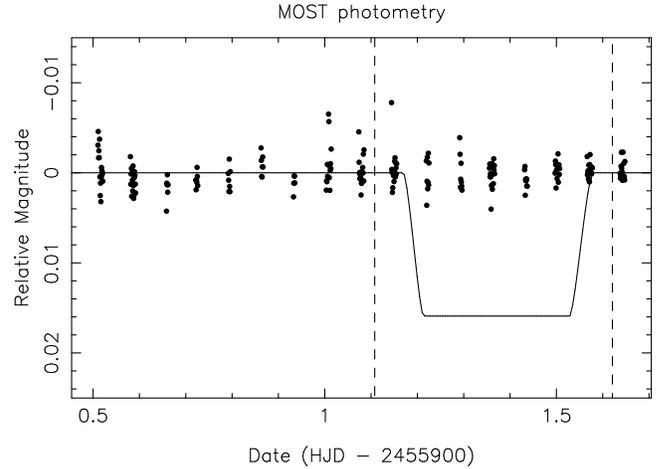}
\caption{Photometric observations of HD 37605 by the MOST satellite,
  which rule out the edge-on transit of HD~37605$b$ at a
  $\gg$ 10$\sigma$ level. The solid line is the predicted transit
  light curve, and the dashed vertical lines are the 1$\sigma$
  transit window boundaries defined by adding $\sigma_{T_c}$ (0.069
  day) on both sides of the predicted transit window
  (0.352-day wide). See \S~\ref{sec:most} for more
  details. \label{most}}
\end{figure}

\renewcommand{\arraystretch}{1.2}
\begin{deluxetable}{cc}
\tabletypesize{\small}
\tablewidth{180pt}
\tablecaption{PHOTOMETRIC OBSERVATIONS OF HD~37605 ON MOST\label{tab:most}}
\tablehead{
\colhead{Heliocentric Julian Date} & \colhead{Relative Magnitude} \\
\colhead{(HJD $-$ 2,451,545)} & \colhead{(mag)}
}
\startdata
     4355.5105   &    -0.0032 \\
     4355.5112   &    -0.0047 \\
     4355.5119   &    -0.0018 \\
     4355.5126   &    -0.0026 \\
     4355.5133   &    -0.0018 \\
     4355.5140   &    -0.0039
\enddata
\tablecomments{This table is presented in its entirety in the electronic edition
of the Astrophysical Journal.  A portion is shown here for guidance regarding
its form and content.}
\end{deluxetable}

As noted earlier, the near-integer period of HD~37605$b$ makes it
difficult to observe from a single longitude. The brightness of the
target and the relatively long predicted transit duration creates
additional challenges for ground-based observations. We thus observed
HD~37605 during 2011 December 5--6 (around the predicted $T_c$ at BJD
$2,455,901.361$ as listed in Table \ref{fittable}) with the MOST
(Microvariability and Oscillations of Stars) satellite launched in
2003 \citep{2003PASP..115.1023W, 2004Natur.430...51M} in the Direct
Imaging mode. This observing technique is similar to ground-based CCD
photometry, allowing to apply traditional aperture and PSF procedures
for data extraction \citep[see e.g.][for
  details]{2006ApJ...646.1241R}. Outlying data points caused by, e.g.,
cosmic rays were removed.

MOST is orbiting with a period of $\sim$ 101 minutes (14.19
cycles per day, cd$^{-1}$), which leads to a periodic artifact induced by the
scattered light from the earthshine. This signal and its harmonics are
further modulated with a frequency of 1 cd$^{-1}$ originating from the
changing albedo of the earth. To correct for this phenomenon, we
constructed a cubic fit between the mean background and the stellar
flux, which was then subtracted from the data. The reduced and
calibrated MOST photometric data are listed in Table \ref{tab:most}.

The MOST photometry is shown in Figure \ref{most} for the transit
window observations. The vertical dashed lines indicate the beginning
and end of the 1~$\sigma$ transit window defined by adding
$\sigma_{T_c}$ (0.069 day) on both sides of the predicted transit
duration of 0.352~days. The solid line shows the predicted transit
model for the previously described planetary parameters.
The mean of the relative photometry is 0.00\% (or 0.00 mag), with a
rms scatter of 0.17\%, and within the predicted transit window there
are 58 MOST observations. Therefore, the standard error on the mean
relative photometry is 0.17\%$/\sqrt{58} = $ 0.022\%. This means that,
for the predicted transit window and a predicted depth of 1.877\%, we
can conclude a null detection of HD 37605$b$'s transit with extremely
high confidence (149$\sigma$).

Note that the above significance is for an edge-on transit with an
impact parameter of $b = 0.0$. A planetary trajectory across the
stellar disk with a higher impact parameter will produce a shorter
transit duration. However, the gap between each cluster of MOST
measurements is 0.06~days which is 17\% of the edge-on transit
duration. In order for the duration to be fit within the data gaps,
the impact parameter would need to be $b > 0.996$. To estimate a more
conservative lower limit for $b$, we now assume the most unfortunate
case where the transit center falls exactly in the middle of one of
the measurement gaps, and also consider the effect of limb darkening
by using the non-linear limb darkening model by
\cite{2002ApJ...580L.171M} with their fitted coefficients for HD
209458. Even under this scenario, we can still conclude the null
detection for any transit with $b<0.951$ at $\gtrsim 5\sigma$ (taking
into account that there are at least $\sim$ 20 observations will fall
within the transit window in this case, though only catching the
shallower parts of the transit light curve).

All of the above is based on the assumption that the planet has the
predicted radius of 1.1 $R_{\rm Jup}$. If in reality the planet is so
small that even a $b=0$ transit would fall below our detection
threshold, it would mean that the planet has a radius of $< 0.36\ R_{\rm
  Jup}$ (a density of $>$ 74.50 g$/$cm$^3$), which seems unlikely.
It is also very unlikely that our MOST photometry has missed the
transit window completely due to an ill-predicted $T_c$. In the
sampling distribution of $T_c$ from BOOTTRAN (with 1000 replicates;
see \S~\ref{sec:fit} and Appendix), there is no $T_c$ that would put
the transit window completely off the MOST coverage. In the
marginalized posterior distribution of $T_c$ calculated via MCMC (see
\S~\ref{SECN:MCMC} and Fig.~\ref{mcmcplot}), there is only 1 such
$T_c$ out of $3 \times 10^4$ (0.003\%).

\section{Summary and Conclusion}\label{sec:summary}

In this paper, we report the discovery of HD 37605$c$ and the
dispositive null detection of non-grazing transits of HD 37605$b$, the first
planet discovered by HET. HD 37605$c$ is the outer planet of the
system with a period of $\sim 7.5$ years on a nearly circular orbit
($e=0.013$) at $a=3.814$ AU. It is a ``Jupiter analog'' with $\msini =
3.366\ \mjup$, which adds one more sample to the currently still small
inventory of such planets (only 10 including HD 37605$c$; see \S
\ref{sec:intro}). The discovery and characterization of ``Jupiter
analogs'' will help understanding the formation of gas giants as well
as the frequency of true solar system analogs. This discovery is a
testimony to the power of continued observation of planet-bearing
stars.

Using our RV data with nearly 8-year long baseline, we refined the
orbital parameters and transit ephemerides of HD 37605$b$. The
uncertainty on the predicted mid-transit time was constrained down to
0.069 day (at and near $T_c = 2,455,901.361$ in BJD), which is small
compared to the transit duration (0.352 day). In fact, just the
inclusion of the two most recent points in our RV data have reduced
the uncertainty on $T_c$ by over 10\%. We have performed transit
search with APT and the MOST satellite. Because of the near-integer
period of HD~37605$b$ and the longitude of Fairborn Observatory, the
APT photometry was unable to cover the transit window. However, its
excellent photometric precision over five observing seasons enabled us
to rule out the possibility of the RV signal being induced by stellar
activity. The MOST photometric data, on the other hand, were able to
rule out an edge-on transit with a predicted depth of 1.877\% at a
$\gg$ 10$\sigma$ level, with a $5\sigma$ lower limit on the impact
parameter of $b \leq 0.951$. This transit exclusion is a further
demonstration of the TERMS strategy, where follow-up RV observations
help to reduce the uncertainty on transit timing and enable transit
searches.

Our best-fit orbital parameters and errors from RVLIN$+$BOOTTRAN were
found to be consistent with those derived from a Bayesian analysis
using MCMC. Based on the best-fit MCMC systems, we performed dynamic
and TTV analysis on the HD 37605 system. Dynamic analysis shows no
sign of orbital resonance and very minimal planet-planet
interaction. We derived a TTV of $\sim$100 s, which is much smaller
than $\sigma_{T_c}$.

We have also performed a stellar analysis on HD 37605, which shows
that it is a metal rich star ($[{\rm Fe/H}] = 0.336 \pm 0.030$) with a
stellar mass of $M_\star = 1.000 \pm0.017\ \msol$ with a radius of
$R_\star = 0.901 \pm 0.015$. The small variation seen in our
photometric data (amplitude $< 0.003$ mag over the course of four
years) suggests that HD 37605 is consistent as being an old, inactive
star that is probably slowly rotating. We tentatively propose that the
rotation period of the star is $57.67 \pm 0.30$ days, based on a weak
periodic signal seen in our APT photometry.

\section{Note on Previously Published Orbital Fits}\label{sec:correction}

In early 2012, we repaired a minor bug in the BOOTTRAN package, mostly
involving the calculation and error bar estimation of $\msini$. As a
result, the $\msini$ values and their errors for two previously
published systems (three planets) need to be updated. They are: HD
114762$b$ \citep{Kane114762}, HD 168443$b$, and HD 168443$c$
\citep{Pilyavsky2011}. Table \ref{tab:corr} lists the updated $\msini$
and error bars.

One additional system, HD 63454 \citep{Kane63454}, was also analyzed
using BOOTTRAN. However, the mass of HD 63454$b$ is small enough
compared to its host mass and thus was not affected by this change.

\renewcommand{\arraystretch}{1.2}
\begin{deluxetable}{lc}
\tabletypesize{\small}
\tablewidth{180pt}
\tablecaption{Updated $\msini$ and Errors for HD 114762$b$ and HD
  168443$b$, $c$\label{tab:corr}}
\tablehead{
\colhead{Planet} & \colhead{$\msini~\pm$ std.~error ($\mjup$)}  
}
\startdata
~~~HD 114762$b$\tablenotemark{a} & $11.086  \pm  0.067$ \\ 
~~~HD 114762$b$\tablenotemark{b} & $11.069  \pm  0.063$ \\ 
~~~HD 168443$b$ & ~$7.696   \pm  0.015$ \\
~~~HD 168443$c$ & $17.378  \pm  0.044$
\enddata
\tablenotetext{a}{For best orbital fit with RV trend ($\mbox{d}
  v/\mbox{d} t$).}
\tablenotetext{b}{For best orbital fit without RV trend ($\mbox{d}
  v/\mbox{d} t$).}
\end{deluxetable}

\acknowledgements
The authors thank John A. Johnson for providing a copy of his Doppler
code and his help with our incorporation of the code into the HET
pipeline.  The authors also thank Debra Fischer for her assistance in
this regard. We thank Peter Plavchan, Scott Dolim, Charley Noecker,
and Farhan Feroz for useful discussions on the term ``dispositive null".

This work was partially supported by funding from the Center for
Exoplanets and Habitable Worlds, which is supported by the
Pennsylvania State University, the Eberly College of Science, and the
Pennsylvania Space Grant Consortium.

The authors appreciate the significant Keck observing time and
associated funding support from NASA for the study of long period
planets and mulitplanet systems.
J.T.W.\ and S.X.W.\ acknowledge support from NASA Origins of Solar
Systems grant NNX10AI52G.
The work of W.D.C., M.E., and P.J.M.\ was supported by NASA Origins of
Solar Systems Grant NNX09AB30G.
E.B.F. and M.J.P. were supported by NASA Origins of Solar Systems
grant NNX09AB35G.
D.D. is supported by a University of British Columbia Four Year
Fellowship.

The work herein is based on observations obtained at the W. M. Keck
Observatory, which is operated jointly by the University of California
and the California Institute of Technology.  The Keck Observatory was
made possible by the generous financial support of the W.M. Keck
Foundation.  We wish to recognize and acknowledge the very significant
cultural role and reverence that the summit of Mauna Kea has always
had within the indigenous Hawaiian community.  We are most fortunate
to have the opportunity to conduct observations from this mountain.

The Hobby-Eberly Telescope is a joint project of the University of
Texas at Austin, the Pennsylvania State University, Stanford
University, Ludwig Maximillians Universit\"at M\"unchen, and Georg
August Universit\"at G\"ottingen. The HET is named in honor of its
principal benefactors, William P. Hobby and Robert E. Eberly.

This work has made use of the Exoplanet Orbit Database at
exoplanets.org, the Extrasolar Planets Encyclopedia at exoplanet.eu,
and of NASA’s Astrophysics Data System Bibliographic Services.

\end{CJK*}

\bibliography{references}

\begin{appendix}
  
\section*{Uncertainties via Bootstrapping}
The uncertainties listed for the orbital parameter
estimates\footnote{Through out the paper and sometimes in this
  Appendix, we refer to the ``{\it estimates} of the parameters'' (as
  distinguished from the ``true parameters'', which are not known and
  can only be estimated) simply as the ``parameters''.} and transit
mid-time $T_c$ are calculated via bootstrapping
\citep{1981,davison1997bootstrap} using the package BOOTTRAN, which we
have made publicly available (see \S~\ref{sec:fit}). It is designed to
calculate error bars for transit ephemerides and the Keplerian
orbital fit parameters output by the RVLIN
package\citep{2009ApJS..182..205W}, but can also be a stand-alone
package. Thanks to the simple concept of bootstrapping, it is
computationally very time-efficient and easy to use.

The basic idea of bootstrap is to resample based on original data
to create bootstrap samples (multiple data replicates); then for
each bootstrap sample, derive orbital parameters or transit parameters
through orbital fitting and calculation. The ensemble of parameters
obtained in this way yields the approximate sampling distribution for
each estimated parameter. The standard deviation of this sampling
distribution is the standard error for the estimate.

We caution the readers here that there are regimes in which the
``approximate sampling distribution" (a frequentist's concept) is not
an estimate of the posterior probability distribution (a Bayesian
concept), and there are regimes (e.g., when limited sampling affects
the shape of the $\chi^2$ surface) where there are qualitative
differences and the bootstrap method dramatically underestimates
uncertainties \citep[e.g., long-period planets when the observations
  are not yet sufficient to pin down the orbital
  period;][]{Ford2005,Bender2012}. In situations with sufficient RV
data, good phase coverage, a sufficient time span of observations and
a good orbital fit, bootstrap often gives a useful estimate of the
parameter uncertainties. For the data considered in this paper, it
was not obvious that the bootstrap uncertainty estimate would be
accurate, as the time span of observations is only slightly longer than
the orbital period of planet $c$. Nevertheless, we find good agreement
between the uncertainty estimates derived from bootstrap and MCMC
calculations.

The radial velocity data are denoted as $\lbrace \vec{t},\ \vec{v},\ \vec{\sigma}
\rbrace$, where each $t_i$, $v_i$, $\sigma_i$ represents radial
velocity $v_i$ observed at time (BJD) $t_i$ with velocity uncertainty
$\sigma_i$. Extreme outliers should be rejected in order to preserve the
validity of our bootstrap algorithm. We first derive our estimates for
the true orbital parameters from the original RV data via orbital fitting,
using the RVLIN package \citep{2009ApJS..182..205W}: \beq \vec{\beta}
= \mu(\vec{t},\ \vec{v},\ \vec{\sigma}), \eeq where $\vec{\beta}$ is
the best fitted orbital parameters\footnote{As described in \S
  \ref{sec:fit}, this includes the $P$, $T_p$, $K$, $e$, and $\omega$
  for each planet, as well as $\gamma$, $\mbox{d}v/\mbox{d}t$ (if
  applicable), and velocity offsets between instruments/telescopes (if
  applicable) for the system.}. From $\vec{\beta}$, we derive $\lbrace
\vec{t},\ \vec{v}_{best}(\vec{\beta}) \rbrace$, the best-fit model
(here $\vec{t}$ are treated as predictors and thus fixed). Then we can
begin resampling to create bootstrap samples.

Our resampling plan is model-based resampling, where we draw from the
residuals against the best-fit model. For data that come from the
same instrument or telescope, in which case no instrumental offset
needs to be taken into account, we simply draw from all residuals,
$\lbrace \vec{v}-\vec{v}_{best} \rbrace$, with equal probability for
each $(v_i - v_{best,i})$. This new ensemble of residuals, denoted as
${\vec{r^*}}$, is then added to the best-fit model $\vec{v}_{best}$ to
create one bootstrap sample, $\vec{v^*}$ \footnote{We simply use the
  raw residual instead of any form of modified residual, because the
  RV data for any single instrument or telescope are usually close
  enough to homoscedasticity.}. Associated with ${\vec{r^*}}$, the
uncertainties $\vec{\sigma}$ are also re-assigned to $\vec{v^*}$ --
that is, if $v_j - v_{best,j}$ is drawn as $r_k$ and added to $v_k$ to
generate $v^*_k$, then the uncertainty for $v^*_k$ is set to be
$\sigma_j$.

For data that come from multiple instruments or multiple
telescopes, we incorporate our model-based resampling plan to include
stratified sampling. In this case, although data from each instrument
or telescope are close to homoscedastic, the entire set of data are
usually highly heteroscedastic due to stratification in
instrument/telescope radial velocity precision. Therefore, the
resampling process is done by breaking down the data into different
groups, $\lbrace \vec{v_1},\ \vec{v_2},\ \ldots \rbrace$, according to
instrument and/or telescope, and then resample within each
subgroup of data with the algorithm described in last paragraph. The bootstrap
sample is then $\vec{v^*} = \lbrace \vec{v^*}_1,\ \vec{v^*}_2,\ \ldots
\rbrace$.

To construct the approximate sampling distribution of the orbital
parameter estimates $\vec{\beta}$, we compute \beq \vec{\beta^*} =
\mu(\vec{t},\ \vec{v^*},\ \vec{\sigma^*}) \eeq for each bootstrap
sample, $\lbrace \vec{t},\ \vec{v^*},\ \vec{\sigma^*} \rbrace$. The
sampling distribution for each orbital parameter estimate $\beta_i$
can be constructed from the multiple sets of $\vec{\beta^*}$
calculated from multiple bootstrap samples
($\vec{\beta^*}^{(1)},\ \vec{\beta^*}^{(2)},\ \ldots$ from
$\vec{v^*}^{(1)},\ \vec{v^*}^{(2)},\ \ldots$). The standard errors for
$\vec{\beta}$ are simply the standard deviations of the sampling
distributions\footnote{The standard deviation of a sampling
  distribution is estimated in a robust way using the IDL function
  {\it robust\_sigma}, which is written by H. Fruedenreich based on
  the principles of robust estimation outlined in \cite{1983ured.book.....H}.}.

The sampling distribution of the estimated transit mid-time, $T_c$, is
calculated likewise. Here $T_c$ is the transit time for a certain
planet of interest in the system, and is usually specified to be the first
transit after a designated time $T$.  However, the situation is
complicated by the periodic nature of $T_c$. Our approach is to first
calculate, based on the original RV data, $T_{c0}$, the estimated
mid-time of the first transit after time $T_0$ (an arbitrary time
within the RV observation time window of $[\min(\vec{t}),\ \max(\vec{t})]$; $T_{c0}$ 
is also within this window).
Then
\beq
T_c = N\cdot P + T_{c0},
\eeq
where $P$ is the best-estimated period for this planet of interest,
and $N$ is the smallest integer that is larger than $(T - T_{c0})/P$.
Next we compute $T_{c0}^*$ for each bootstrap sample $\lbrace
\vec{t},\ \vec{v^*},\ \vec{\sigma^*} \rbrace$. Given that within the
time window of radial velocity observations
($[\min(\vec{t}),\ \max(\vec{t})]$), the phase of the planet should be
known well enough, it is fair to assume that $T_{c0}$ is an unbiased
estimator of the true transit mid-time. Therefore we assert that
$T_{c0}^*$ has to be well constrained and within the range of
$[T_{c0}-P^*/2,\ T_{c0}+P^*/2]$, where $P^*$ is the period estimated from
this bootstrap sample. If not, then we subtract or add multiple
$P^*$'s until $T_{c0}^*$ falls within the range. Then naturally
\beq
T_c^* = N\cdot P^* + T_{c0}^*.
\eeq
The ensemble of $T_c^*$'s gives the sampling distribution of $T_c$
and its standard error. Note that $T_c^*$ is not necessarily within
the rage of $[T_{c}-P/2,\ T_{c}+P/2]$.

Provided with the stellar mass $M_\star$ and its uncertainty, we
calculate, for each planet in the system, the standard errors for the
semi-major axis $a$ and the {\it minimum mass} of the planet $M_{\rm
  p,min}$ (denoted as $\msini$ in the main text as commonly seen in
literature, but this is a somewhat imprecise notation). As the first
step, the mass function is calculated for the best-fit $\vec{\beta}$
and each bootstrap sample $\vec{\beta^*}$,
\beq
f(P,K,e) = \frac{P K^3
  (1-e)^{3/2}}{2\pi G} = \frac{(M_p\cdot \sin i)^3}{(M_\star+M_p)^2}.
\eeq
The sampling distribution of $f(P,K,e)$ then gives the standard error
of the mass function. The minimum mass of the planet $M_{\rm p,min}$ 
is then calculated by assuming $\sin{i}=1$ and solving for $M_p$.
Standard error of $M_{\rm p,min}$ is derived through simple
propagation of error, as the covariance between $M_\star$ and
$f(P,K,e)$ is probably negligible.

For the semi-major axis $a$,
\beq
a^3 = \frac{P^2 G
  (M_\star+M_p)}{4\pi^2} \approx \frac{P^2 G (M_\star + M_{\rm p,min})}{4\pi^2}.
\eeq
The standard error of $P^2$ is calculated from its bootstrap sampling
distribution, and via simple propagation of error we obtain the
standard error of $a$ (neglecting covariance between $P^2$, $M_{\rm
  p,min}$, and $M_\star$).

\end{appendix}

\clearpage
\LongTables 
\renewcommand{\arraystretch}{1.2} 
\begin{deluxetable}{lrrc}
\tabletypesize{\scriptsize}
\tablecaption{HET and KECK RADIAL VELOCITIES FOR HD 37605\label{rvtable}}
\tablewidth{0pt}
\tablehead{
\colhead{} & \colhead{Velocity} & \colhead{Uncertainty} & \colhead{} \\
\colhead{~BJD$-2440000\ \ $} & \colhead{(\mps)} &
\colhead{(\mps)} & \colhead{Telescope} 
}
\startdata
        13002.671503 &    122.4 &   8.8 & HET \\
        13003.685247 &    126.9 &   5.6 & HET \\
        13006.662040 &    132.1 &   5.2 & HET \\
        13008.664059 &    130.4 &   4.8 & HET \\
        13010.804767 &    113.4 &   4.7 & HET \\
        13013.793987 &    106.2 &   4.9 & HET \\
        13042.727963 &   -120.0 &   4.7 & HET \\
        13061.667551 &    126.5 &   4.0 & HET \\
        13065.646834 &    111.3 &   5.8 & HET \\
        13071.643820 &    106.7 &   5.5 & HET \\
        13073.638180 &     91.7 &   4.6 & HET \\
        13082.623709 &     53.4 &   5.7 & HET \\
        13083.595357 &     45.1 &   6.4 & HET \\
        13088.593775 &     30.7 &  11.3 & HET \\
        13089.595750 &      0.8 &   5.5 & HET \\
        13092.597983 &    -25.5 &   6.2 & HET \\
        13094.586570 &    -51.0 &   5.9 & HET \\
        13095.586413 &    -72.2 &   6.2 & HET \\
        13096.587432 &    -75.0 &   9.3 & HET \\
        13098.576213 &   -173.1 &   6.4 & HET \\
        13264.951365 &   -194.0 &   9.8 & HET \\
        13265.947438 &   -238.2 &  10.5 & HET \\
        13266.945977 &   -264.2 &  13.1 & HET \\
        13266.959470 &   -286.8 &  11.6 & HET \\
        13283.922407 &    118.8 &   9.6 & HET \\
        13318.819260 &   -142.8 &   6.6 & HET \\
        13335.921770 &    143.4 &   6.3 & HET \\
        13338.906008 &    132.1 &   5.4 & HET \\
        13377.819405 &   -279.6 &   6.1 & HET \\
        13378.811880 &   -161.3 &   5.2 & HET \\
        13379.802247 &    -39.7 &   5.2 & HET \\
        13381.644284 &     74.4 &   4.7 & HET \\
        13384.646538 &    112.5 &   5.8 & HET \\
        13724.855831 &     88.1 &   5.3 & HET \\
        13731.697227 &     57.8 &   5.0 & HET \\
        13738.674709 &     36.4 &   4.9 & HET \\
        13743.810196 &     14.1 &   5.7 & HET \\
        13748.647234 &    -17.8 &   5.6 & HET \\
        14039.850147 &   -104.1 &   5.5 & HET \\
        14054.964568 &     68.8 &   7.3 & HET \\
        14055.952778 &     49.3 &   6.5 & HET \\
        14067.762810 &     22.1 &   6.0 & HET \\
        14374.924086 &     32.7 &   8.6 & HET \\
        14394.864468 &    -12.1 &   8.5 & HET \\
        14440.883298 &     11.9 &   8.5 & HET \\
        14467.826532 &   -121.9 &   6.3 & HET \\
        14487.613032 &     40.4 &   5.6 & HET \\
        14515.689341 &    -55.6 &   5.7 & HET \\
        14550.600301 &     23.1 &   5.5 & HET \\
        14759.878895 &     34.3 &   4.9 & HET \\
        14776.839187 &      2.0 &   6.7 & HET \\
        14787.793398 &    -40.8 &   5.1 & HET \\
        14907.617522 &    -94.4 &   5.1 & HET \\
        15089.968860 &     76.4 &   4.1 & HET \\
        15104.917965 &     54.0 &   4.6 & HET \\
        15123.882222 &    -44.4 &   4.6 & HET \\
        15172.734743 &     -0.4 &   4.6 & HET \\
        15172.749345 &      3.0 &   4.9 & HET \\
        15180.715307 &    -65.0 &   5.3 & HET \\
        15190.692697 &   -257.0 &   4.9 & HET \\
        15190.704818 &   -252.3 &   6.3 & HET \\
        15195.678923 &    -47.7 &   5.1 & HET \\
        15200.809909 &     85.1 &   4.5 & HET \\
        15206.665616 &     75.9 &   4.5 & HET \\
        15208.646672 &     72.6 &   5.7 & HET \\
        15266.634586 &     81.1 &   4.6 & HET \\
        15274.598361 &     57.1 &   4.6 & HET \\
        15280.605398 &     24.3 &   5.6 & HET \\
        15450.974863 &     25.2 &   5.2 & HET \\
        15469.928975 &    -81.7 &   4.4 & HET \\
        15481.884352 &    110.8 &   4.8 & HET \\
        15494.004309 &     71.2 &   4.9 & HET \\
        15500.853302 &     41.1 &   6.0 & HET \\
        15505.824457 &     21.2 &   4.8 & HET \\
        15510.809500 &    -13.9 &   5.5 & HET \\
        15511.813809 &    -18.4 &   4.7 & HET \\
        15518.798148 &   -139.1 &   4.9 & HET \\
        15518.943518 &   -148.8 &   4.6 & HET \\
        15524.791579 &    -97.7 &   6.0 & HET \\
        15526.759230 &     51.2 &   5.8 & HET \\
        15526.767959 &     56.5 &   6.1 & HET \\
        15527.921902 &     89.2 &   4.6 & HET \\
        15528.921753 &    105.1 &   5.5 & HET \\
        15530.757847 &    120.6 &   5.3 & HET \\
        15532.760658 &    117.6 &   4.8 & HET \\
        15544.730827 &     81.8 &   4.9 & HET \\
        15545.871710 &     87.7 &   5.2 & HET \\
        15547.717980 &     78.6 &   6.5 & HET \\
        15547.720596 &     81.2 &   7.8 & HET \\
        15550.703130 &     75.8 &   5.5 & HET \\
        15556.688398 &     43.4 &   5.0 & HET \\
        15557.842853 &     40.1 &   4.9 & HET \\
        15566.657803 &    -18.4 &   5.6 & HET \\
        15566.666973 &    -19.0 &   4.9 & HET \\
        15569.670948 &    -50.7 &   4.5 & HET \\
        15571.659560 &    -94.0 &   7.4 & HET \\
        15571.662188 &    -99.5 &   7.5 & HET \\
        15582.777052 &     75.1 &   5.4 & HET \\
        13982.116400 &   -312.6 &   1.1 & Keck \\
        13983.110185 &   -291.6 &   1.1 & Keck \\
        13984.104595 &   -175.0 &   1.1 & Keck \\
        14024.076794 &    -39.9 &   1.2 & Keck \\
        14129.848461 &    -17.1 &   1.1 & Keck \\
        14131.926354 &    -32.2 &   1.3 & Keck \\
        14138.913472 &    -98.8 &   1.2 & Keck \\
        14544.815162 &     54.2 &   1.0 & Keck \\
        14545.799155 &     50.4 &   1.0 & Keck \\
        14546.802812 &     48.1 &   1.0 & Keck \\
        15172.939733 &     21.7 &   1.1 & Keck \\
        15435.133695 &     97.5 &   1.0 & Keck \\
        15521.851677 &   -244.2 &   1.2 & Keck \\
        15522.947381 &   -268.1 &   1.1 & Keck \\
        15526.936103 &     78.6 &   1.1 & Keck \\
        15528.877669 &    118.4 &   1.0 & Keck \\
        15543.032113 &    115.5 &   1.2 & Keck \\
        15545.105481 &    110.3 &   1.2 & Keck \\
        15546.046328 &    110.6 &   1.2 & Keck \\
        15556.071287 &     69.6 &   1.1 & Keck \\
        15556.954989 &     66.4 &   1.1 & Keck \\
        15558.027806 &     62.3 &   1.2 & Keck \\
        15584.769501 &    128.2 &   1.2 & Keck \\
        15606.952224 &     91.6 &   1.2 & Keck \\
        15634.789262 &    -81.0 &   1.0 & Keck \\
        15636.793990 &     80.7 &   1.0 & Keck \\
        15636.799360 &     85.6 &   1.1 & Keck \\
        15663.738831 &     87.6 &   1.0 & Keck \\
        15671.731754 &     45.8 &   1.2 & Keck \\
        15672.732058 &     37.1 &   1.0 & Keck \\
        15673.745070 &     27.6 &   1.0 & Keck \\
        15793.124325 &    -86.7 &   1.2 & Keck \\
        15843.017282 &     -6.4 &   1.3 & Keck 
\enddata
\end{deluxetable}

\end{document}